\documentclass[aps,prc,twocolumn,10pt,superscriptaddress]{revtex4-1}

\usepackage[utf8]{inputenc}
\usepackage[margin=0.5in]{geometry}
\usepackage{hyperref}
\usepackage{amsmath,amsfonts,amssymb}
\usepackage{graphicx}
\usepackage[usenames,dvipsnames,svgnames,table]{xcolor}
\usepackage[normalem]{ulem} 

\hypersetup{
    colorlinks,
    unicode=True,
    linkcolor=Blue,
    citecolor=Blue,
    urlcolor=Blue,
    pdftitle={Jet transport coefficients by elastic and radiative scatterings in the strongly interacting quark-gluon plasma},
    pdfauthor={Ilia Grishmanovskii},
}

\graphicspath{ {figures/} }

\newcommand{\dpi}{(2\pi)}
\newcommand{\qhat}{\hat{q}}
\newcommand{\calO}{\mathcal{O}}
\renewcommand{\vec}{\mathbf}

\begin{document}


\title{Jet transport coefficients by elastic and radiative scatterings in the strongly interacting quark-gluon plasma}


\author{Ilia Grishmanovskii}
\email{grishm@itp.uni-frankfurt.de}
\affiliation{Institut f\"ur Theoretische Physik, Johann Wolfgang Goethe-Universit\"at,Max-von-Laue-Straße 1, D-60438 Frankfurt am Main, Germany}

\author{Olga Soloveva}
\affiliation{Helmholtz Research Academy Hesse for FAIR (HFHF), GSI Helmholtz Center for Heavy Ion Physics, Campus Frankfurt, 60438 Frankfurt, Germany}
\affiliation{Institut f\"ur Theoretische Physik, Johann Wolfgang Goethe-Universit\"at,Max-von-Laue-Straße 1, D-60438 Frankfurt am Main, Germany}

\author{Taesoo Song}
\affiliation{GSI Helmholtzzentrum f\"ur Schwerionenforschung GmbH, Planckstrasse 1, D-64291 Darmstadt, Germany}

\author{Carsten Greiner}
\affiliation{Institut f\"ur Theoretische Physik, Johann Wolfgang Goethe-Universit\"at,Max-von-Laue-Straße 1, D-60438 Frankfurt am Main, Germany}
\affiliation{Helmholtz Research Academy Hesse for FAIR (HFHF), GSI Helmholtz Center for Heavy Ion Physics, Campus Frankfurt, 60438 Frankfurt, Germany}

\author{Elena Bratkovskaya}
\affiliation{GSI Helmholtzzentrum f\"ur Schwerionenforschung GmbH, Planckstrasse 1, D-64291 Darmstadt, Germany}
\affiliation{Institut f\"ur Theoretische Physik, Johann Wolfgang Goethe-Universit\"at,Max-von-Laue-Straße 1, D-60438 Frankfurt am Main, Germany}
\affiliation{Helmholtz Research Academy Hesse for FAIR (HFHF), GSI Helmholtz Center for Heavy Ion Physics, Campus Frankfurt, 60438 Frankfurt, Germany}

\date{\today}


\begin{abstract}
We extend the investigation of jet transport coefficients within the effective Dynamical QuasiParticle Model (DQPM) -- constructed to describe nonperturbative QCD phenomena of the strongly interacting quark-gluon plasma (sQGP) in line with the lattice QCD equation of state -- by accounting for inelastic $2\to 3$ reactions with gluon radiation in addition to the elastic scattering of partons. The elastic and inelastic reactions are calculated explicitly within leading-order Feynman diagrams with effective propagators and vertices from the DQPM by accounting for all channels and their interferences. We present the results for the jet transport coefficients such as the transverse momentum transfer squared $\hat{q}$ per unit length as well as the energy loss $dE/dx$ per unit length in the sQGP and investigate their dependence on the temperature $T$ and momentum of the jet parton depending on the choice of the strong coupling constant $\alpha_s$ in thermal, jet parton and radiative vertices. For the latter, we consider different scenarios used in the literature and find a very strong dependence of $\qhat$ and $dE/dx$ on the choice of $\alpha_s$. Moreover, we explore the relation of $\hat{q}/T^3$ to the ratio of specific shear viscosity to entropy density $\eta/s$ and show that the ratio $T^3/\hat{q}$ to $\eta/s$ has a strong $T$ dependence -- especially when approaching $T_c$ -- on the choice of $\alpha_s$ in scattering vertices.
\end{abstract}

\maketitle


\section{Introduction}

Jets are considered as one of the penetrating probes that allow to obtain information about the strong interactions in the QGP medium created in heavy-ion collisions. As observed experimentally at RHIC \cite{STAR:2003pjh,STAR:2002svs} and at LHC \cite{ALICE:2010yje,ATLAS:2010isq}, jets produced in these A+A collisions are modified compared to those produced in proton-proton (p+p) collisions due to the scattering of partons from the jet shower with the partons from the QGP environment. Theoretical understanding of the experimental data on jet attenuation in heavy-ion collisions stimulated a lot of theoretical efforts on an understanding of jet properties in both equilibrium and non-equilibrium cases \cite{Thoma:1990fm,Dokshitzer:1991fc,Mrowczynski:1991da,Gyulassy:1993hr, Baier:1996kr, Baier:1996sk, Baier:1998kq, Zakharov:1996fv, Gyulassy:1999zd, Wiedemann:2000za, Guo:2000nz, Arnold:2002ja, Armesto:2003jh,Moore:2004tg, Djordjevic:2008iz, Zapp:2008gi,Majumder:2009ge,Caron-Huot:2010qjx, Stojku:2020tuk, Barata:2020sav,Cao:2020wlm}. The jet energy loss in the QGP medium occurs due to elastic $2\to 2$ partonic scatterings as well as by radiative $2\to 3$ processes with the emission of a gluon. The soft gluon radiation is screened in the medium due to the coherence effect called Landau-Pomeranchuk-Migdal (LPM) effect \cite{Landau:1953um,Migdal:1956tc}, which has to be accounted for in the interpretation of experimental data. The first calculations of the gluon Bremsstrahlung processes with the non-Abelian LPM effect in the context of perturbative QCD \cite{Baier:1994bd,Baier:1996vi,Wang:1994fx,Zakharov:1996fv,Zakharov:2020whb} showed a suppression of soft gluon radiation due to the destructive interference of the emitted gluons in the medium, contrary to the vacuum; cf. also Refs. \cite{Gyulassy:1999zd,Wiedemann:2000za,Arnold:2008iy,Zapp:2011ya,Armesto:2011ht,Casalderrey-Solana:2011ule,Ke:2018jem,Senzel:2020tgf}. A further formulation of radiative energy loss -- as a dynamical process in the QGP medium of the thermal scattering centers -- has been derived by Arnold, Moore, and Yaffe (AMY) within thermal field theory \cite{Arnold:2002zm,Arnold:2002ja,Arnold:2000dr} and developed further in recent years by going beyond the limitations of soft-emission approximations \cite{Arnold:2020uzm,Arnold:2021pin}. Also, progress has been achieved in the resummation of multiple scatterings \cite{Andres:2022ndd,Andres:2023jao,Barata:2020sav}.

Since the jets are multi-scale objects, the microscopic simulation of jet propagation and interactions with the QCD medium is notoriously difficult. The description of partonic energy loss in the QGP is widely studied in the literature in terms of transport coefficients, such as the transverse momentum transfer squared $\qhat$ per unit length to characterize the jet transverse momentum broadening in the medium, as well as the energy loss $dE/dx$ per unit length to account for the modification of jet energy during propagation in the medium.

The $\qhat$ represents a measure of the interaction between a high-energetic jet and the QGP medium, which is defined as the amount of momentum transfer that the hard parton experiences per unit length as it travels through the dense QGP medium. The value of this coefficient depends on several factors, including the coupling strength of the interaction, the nature of the plasma (whether it is dominated by quasiparticles or not), and the underlying micro-physics of many-body QCD matter.

Recently, in Ref. \cite{Grishmanovskii:2022tpb} the jet transport coefficients $\qhat$ and $dE/dx$ have been studied for elastic $2\to 2$ interactions of partons within the effective dynamical quasiparticle model (DQPM), constructed to describe the nonperturbative properties of the sQGP at finite temperature $T$ and baryon chemical potential $\mu_B$ in terms of strongly interacting off-shell partons (quarks and gluons) with dynamically generated spectral functions, whose properties are adjusted to reproduce the lQCD EoS for the QGP in thermodynamic equilibrium. It has been shown that $\qhat$ and $dE/dx$ show a strong temperature dependence due to the growing strong coupling $\alpha_s(T)$ when approaching the critical temperature $T_c$. However, additionally to elastic scattering, the gluon radiative processes $2\to 3$ are also important for an understanding of the high energetic jet attenuation in heavy-ion collisions, whose contribution grows with increasing energy of jet partons -- cf. Refs. \cite{Djordjevic:2006tw,Stojku:2020tuk,Cao:2020wlm} and references therein.

In Ref. \cite{Grishmanovskii:2023gog} the massive gluon radiation processes from the off-shell quark-quark ($q+q$) and quark-gluon ($q+g$) scatterings have been calculated explicitly (for the first time) within the DQPM based on leading order Feynman diagrams with effective propagators and vertices from the DQPM without any further approximations. There the total and differential radiative cross sections have been evaluated versus the collision invariant energy $\sqrt{s}$, temperature $T$, and baryon chemical potential $\mu_B$ and compared to the corresponding elastic cross sections. While in the limit of zero masses and widths of quasiparticles, the DQPM reproduces the results of pQCD for $2 \to 3$ cross sections \cite{Grishmanovskii:2023gog}, the results for the radiation of heavy gluons from quasiparticle scattering may differ substantially from the pQCD calculations. Thus, one expects an influence of the radiative process on the interaction of fast partons propagated through the equilibrated sQGP medium of quasiparticles.

In the present work, we aim to understand how important gluon radiative reactions are for the jet transport coefficient in the case of massive quarks and gluons and explore the temperature and momentum dependence of the transport coefficients. The latter can be of great interest in the case of jet quenching models, such as JEWEL \cite{Zapp:2009ud}, JET \cite{Burke:2013yra}, JETSCAPE \cite{JETSCAPE:2021ehl}, and etc. It is important to note that we do not use the assumption of perturbative QCD (pQCD), which states that the interaction between an energetic parton and the medium is dominated by small-angle scattering and induced gluon radiation. In this work, we aim to evaluate jet transport coefficients by properly taking into account all diagrams and channels involved in $2\to 2$ and $2 \to 3$ partonic scattering with massive medium partons as described by DQPM, i.e., the in-medium modification of jet partons occurs by sequences of independent elastic and inelastic reactions with thermal partons from the medium at given temperature $T$.

The framework of quasiparticle models facilitates its implementation in the transport model for the dynamical evolution of the QGP matter. In particular, the off-shell quasiparticle model DQPM is implemented in the Parton-Hadron-String Dynamics (PHSD) transport approach \cite{Cassing:2009vt,Moreau:2019vhw}, whereas the on-shell QPM \cite{Plumari:2011mk} is implemented in the Catania transport approach \cite{Scardina:2017ipo}. 

In this study, we also investigate the dependence of elastic and inelastic transport coefficients on the choice of the strong coupling constant $\alpha_s$ used in thermal, jet parton and gluon radiative vertices, which could differ from the coupling constant for the thermalized partons, since the jet parton is not in equilibrium with the QGP medium \cite{Zakharov:2023oav, Boguslavski:2023waw,Shi:2022rja}. For that goal we select four models for $\alpha_s$ -- used in the literature -- and compare the results with the default DQPM $\alpha_s(T)$. Also we explore the dependence of elastic and inelastic $\qhat$ and $dE/dx$ on the mass of the emitted gluon. Some model cases we compare to the results from the Boltzmann Approach to Multi-Parton Scatterings (BAMPS) pQCD cascade, which explicitly includes $2\to 2$ and $2\to 3$ (as well as backward) reactions \cite{Uphoff:2014cba,Fochler:2013epa} as well as the LPM effect \cite{Senzel:2020tgf}.

Furthermore, we investigate the relation of $\qhat/T^3$ and the specific shear viscosity $\eta/s$ discussed in \cite{Muller:2021wri}.

This paper is organized as follows. In Sec. \ref{sec:DQPM} we recall the main ideas of the DQPM, in Sec. \ref{sec:methodology} we describe the framework for the calculation of transport coefficients and present the scenarios for the coupling constant. In Secs. \ref{sec:res_dxs}, \ref{sec:res_qhat}, and \ref{sec:eta-qhat} we report on the results for the radiative differential cross sections and transport coefficients. We summarize our study in Sec. \ref{sec:conclusion}.


\section{Dynamical Quasiparticle Model}
\label{sec:DQPM}

The Dynamical Quasiparticle Model (DQPM) \cite{Peshier:2005pp,Cassing:2007nb,Cassing:2007yg,Berrehrah:2016vzw,Moreau:2019vhw,Soloveva:2020hpr} is an effective model that describes the Quark-Gluon Plasma (QGP) in terms of strongly interacting quarks and gluons. This approach involves fitting the properties of these quasiparticles to match the results of lattice QCD calculations in thermal equilibrium and at vanishing chemical potential. Here we briefly recall the basic ideas of the DQPM. The quasiparticles in the DQPM are characterized by "dressed" propagators, i.e., single-particle (two-point) Green's functions, which have the form
\begin{equation}
    G^{R}_j (\omega, \vec{p}) = \frac{1}{\omega^2 - \vec{p}^2 - M_j^2 + 2 i \gamma_j \omega}
    \label{eq:propdqpm}
\end{equation}
for quarks, antiquarks, and gluons ($j = q,\bar q,g$), using $\omega=p_0$ for energy, the widths $\gamma_{j}$ and the masses $M_{j}$, and the complex self-energies for gluons $\Pi = M_g^2-2i \omega \gamma_g$ and for (anti)quarks $\Sigma_{q} = M_{q}^2 - 2 i \omega \gamma_{q}$, where the real part of the self-energies is associated with dynamically generated thermal masses, while the imaginary part provides information about the lifetime and reaction rates of the particles.

The spectral function of off-shell quasiparticles in the DQPM are parametrized in Lorenzian form with a finite width $\gamma_{j}$ \cite{Linnyk:2015rco}:
\begin{align}
    \rho_{j}(\omega,\vec{p}) &= \frac{\gamma_{j}}{\tilde{E}_j}
    \left(\frac{1}{(\omega-\tilde{E}_j)^2+\gamma^{2}_{j}}
    -\frac{1}{(\omega+\tilde{E}_j)^2+\gamma^{2}_{j}}\right) 
    \nonumber\\
    &\equiv \frac{4\omega\gamma_j}{\left( \omega^2 - \vec{p}^2 - M^2_j \right)^2 + 4\gamma^2_j \omega^2},
    \label{eq:spectral_function}
\end{align}
with $\tilde{E}_{j}^2(\vec{p})=\vec{p}^2+M_{j}^{2}-\gamma_{j}^{2}$. The spectral function is antisymmetric in $\omega$ and normalized as
\begin{equation}
    \int\limits_{-\infty}^{\infty}\frac{d\omega}{2\pi}\
    \omega \ \rho_{j}(\omega,\vec{p})=
    \int\limits_{0}^{\infty} d\omega \frac{\omega}{\pi}\ 
    \rho_{j}(\omega,\vec{p})=1.
    \label{eq:spectral_function_norm}
\end{equation}

The DQPM involves a model ansatz for the masses $M_{j}(T,\mu_q)$ and widths $\gamma_{j}(T,\mu_q)$ as functions of the temperature $T$ and the quark chemical potential $\mu_q$. By comparison of the entropy density -- computed within the DQPM framework -- to the lQCD data, one can fix the few parameters used in the ansatz for the quasiparticle masses and widths. The following ansatz is used in the DQPM for the definition of the quasiparticle properties (masses and widths) as functions of $T$ and $\mu_q$. The dynamical quasiparticle \textbf{pole masses} are given by the HTL thermal mass in the asymptotic high-temperature regime -- cf. \cite{Bellac:2011kqa,Linnyk:2015rco}
\begin{equation}
    M^2_{g}(T,\mu_q)=\frac{g^2(T,\mu_q)}{6}\left(\left(N_{c}+\frac{1}{2}N_{f}\right)T^2
    +\frac{N_c}{2}\sum_{q}\frac{\mu^{2}_{q}}{\pi^2}\right),
    \label{eq:Mg}
\end{equation}
and for quarks (antiquarks) by
\begin{equation}
    M^2_{q(\bar q)}(T,\mu_q)=\frac{N^{2}_{c}-1}{8N_{c}}g^2(T,\mu_q)\left(T^2+\frac{\mu^{2}_{q}}{\pi^2}\right),
    \label{eq:Mq}
\end{equation}
where $N_{c}\ (=3)$ stands for the number of colors and $N_{f}\ (=3)$ denotes the number of light flavors. Equation \eqref{eq:Mq} determines the pole masses for the ($u,d$) quarks; the strange quark has a larger bare mass for controlling the strangeness ratio in the QGP. Empirically, we find $M_s(T,\mu_B)= M_{u/d}(T,\mu_B)+ \Delta M$, where $\Delta M \simeq 30$ MeV has been fixed once in comparison to experimental data \cite{Moreau:2019vhw}. 

The effective quarks, antiquarks, and gluons in the DQPM acquire sizable \textbf{widths} $\gamma_j$, which are taken in the form \cite{Linnyk:2015rco}
\begin{equation}
    \gamma_{j}(T,\mu_\mathrm{B}) = \frac{1}{3} C_j \frac{g^2(T,\mu_\mathrm{B})T}{8\pi}\ln\left(\frac{2c_m}{g^2(T,\mu_\mathrm{B})}+1\right).
    \label{eq:widths}
\end{equation}
Here $c_m=14.4$ is related to a magnetic cutoff, which is an additional parameter in the DQPM, while $C_q = \dfrac{N_c^2 - 1}{2 N_c} = 4/3$ and $C_g = N_c = 3$ are the QCD color factors for quarks and gluons, respectively. We also assume that all (anti)quarks have the same $T$ dependence for the width. 

The coupling constant $g^2 = 4\pi\alpha_s$ essentially defines the masses and widths of quasiparticles -- cf. Eqs. \eqref{eq:Mq}, \eqref{eq:widths}. The thermal properties of quasiparticles and their interactions (observed via transport coefficients) thus strongly depend on the coupling constant $g$.

In the DQPM, $g^2$ is extracted from lattice Quantum Chromodynamics (lQCD) data on the entropy density $s$ by a parametrization method introduced in Ref. \cite{Berrehrah:2015vhe}. There it has been found that for a given value of $g^2$, the ratio $s(T,g^2)/T^3$ is almost constant for different temperatures, i.e., ${\frac{\partial}{\partial T}} (s(T,g^2)/T^3)=0$. Therefore, the entropy density $s$ and the dimensionless equation of state in the DQPM is a function of the effective coupling only, i.e., $s(T,g^2)/s_{SB}(T) = f(g^2)$, where $s^\mathrm{QCD}_{SB} = 19/9 \pi^2T^3$ is the Stefan-Boltzmann entropy density. Thus, by inverting the $f(g^2)$ function, the coupling constant $g^2$ can be directly obtained from the parametrization of lQCD data for the entropy density $s(T,\mu_B=0)$ at zero baryon chemical potential:
\begin{equation}
    g^2(T,\mu_B=0) = d \left( (s(T,0)/s^\mathrm{QCD}_{SB})^e - 1 \right)^f.
    \label{eq:coupling_DQPM}
\end{equation}
Here $d = 169.934, e = -0.178434$, and $f = 1.14631$ are the dimensionless parameters obtained by adjusting the quasiparticle entropy density $s(T,\mu_B=0)$ to the lQCD data provided by the BMW Collaboration \cite{Borsanyi:2012cr, Borsanyi:2013bia}. 

The DQPM $\alpha_s$ accounts for nonperturbative effects and is larger compared to the analytical two- or one-loop running constant \cite{Caswell:1974gg} when approaching low temperatures.

The extension of the coupling constant to finite baryon chemical potential $\mu_B$ is realized using a scaling hypothesis \cite{Cassing:2008nn}, which works up to $\mu_B \approx 500$ MeV. 

Thus, the DQPM provides the quasiparticle properties, dressed propagators, and coupling constant, which can be used to evaluate the scattering amplitudes as well as the cross sections and the transport coefficients of quarks and gluons in the QGP as a function of the temperature and the chemical potential -- cf. Refs. \cite{Berrehrah:2013mua,Moreau:2019vhw,Grishmanovskii:2023gog}.


\section{Methodology}
\label{sec:methodology}

The propagation of a fast jet parton through the thermalized medium can be characterized by jet transport coefficients that can be calculated within kinetic transport theory by accounting for the sequence of elastic and inelastic interactions of the jet parton per unit length $x$ (or time, using $x=ct$, where $c$ is the speed of light; we use the units $c=1$ throughout this work). We note that in this study we do not consider a suppression of the radiation of soft gluons in finite medium due to the LPM effects.

We start with the general expression for a transport coefficient in kinetic theory \cite{PhysRevD.37.2484, PhysRevC.57.889, Moore:2004tg, Berrehrah:2014kba} for \textit{elastic $2\to 2$ reactions}:
\begin{align}
    \langle \calO \rangle^{\text{el}} = \frac{1}{2E_{\text{jet}}}\sum_{i=q,\bar{q},g}
    \int\frac{d^3p_i}{\dpi^3 2E_i} d_i f_i \int\frac{d^3p_1}{\dpi^3 2E_1}
    \nonumber\\
    \times \int\frac{d^3p_2}{\dpi^3 2E_2} (1 \pm f_1)(1 \pm f_2)
    \nonumber\\
    \times \ \calO \ |\bar M_{2\to 2}|^2_{\text{jet} + i} \ (2\pi)^4 \delta^{(4)}(p_{\text{jet}} + p_i - p_1 - p_2),
    \label{eq:O_el}
\end{align}
where $p_i$ is the 4-momentum of the incoming medium parton, $p_1$ and $p_2$ are the outgoing jet and medium parton 4-momenta, respectively; $d_i$ is the medium parton's degeneracy factor for spin and color ($2N_c$ for quarks and $2(N_c^2-1)$ for gluons); $f_i = f_i(E_i,T,\mu_q)$ are the Fermi distribution functions for quarks, and $f_i = f_i(E_i,T)$ are the Bose distribution functions for gluons.

In case of \textit{the inelastic $2 \to 3$ reaction} the expression for the jet transport coefficients takes the following form:
\begin{align}
    \langle \calO \rangle^{\text{inel}} = \frac{1}{2E_{\text{jet}}}\sum_{i=q,\bar{q},g}
    \int\frac{d^3p_i}{\dpi^3 2E_i} d_i f_i \int\frac{d^3p_1}{\dpi^3 2E_1}
    \nonumber\\
    \times \int\frac{d^3p_2}{\dpi^3 2E_2} \int\frac{d^3p_3}{\dpi^3 2E_3} 
    (1 \pm f_1)(1 \pm f_2)(1 \pm f_3)
    \nonumber\\
    \times \ \calO \  |\bar M_{2\to 3}|^2_{\text{jet} + i} \ (2\pi)^4 \delta^{(4)}(p_{\text{jet}} +  p_i - p_1 - p_2 - p_3),
    \label{eq:O_inel}
\end{align}
where $p_3$ denotes the momentum of the emitted gluon.

Various \textit{transport coefficients} can be calculated by selecting different operators $\calO$ in equations \eqref{eq:O_el} and \eqref{eq:O_inel}:
\begin{itemize}
    \item $\calO = 1$ -- scattering rate $\Gamma$,
    \item $\calO = |p_T - p_T^{\prime}|^2$ -- jet transport coefficient $\qhat$,
    \item $\calO = E - E'$ -- energy loss $dE/dx$ per unit length,
    \item $\calO = p_L - p_L'$ -- drag coefficient $\mathcal{A}$.
\end{itemize}
Here $E, E^\prime$, $p_T, p_T^\prime$, and $p_L, p_L^\prime$ denote initial and final energy, transverse and longitudinal momenta of the jet parton, respectively.

In Eq. (\ref{eq:O_inel}) the matrix elements squared, i.e., $|\bar M_{2\to 2}|^2_{\text{jet} + i}$ for elastic $2\to 2$ parton scattering and $|\bar M_{2\to 3}|^2_{\text{jet} + i}$ for inelastic $2 \to 3$ reactions with gluon radiation from the off-shell quark-quark ($q+q$) and quark-gluon ($q+g$) scatterings, are calculated explicitly within the DQPM based on leading order Feynman diagrams with the effective propagators and vertices from the DQPM without any further approximations.

\begin{figure*}[ht!]
    \centering
    \includegraphics[width=0.48\textwidth]{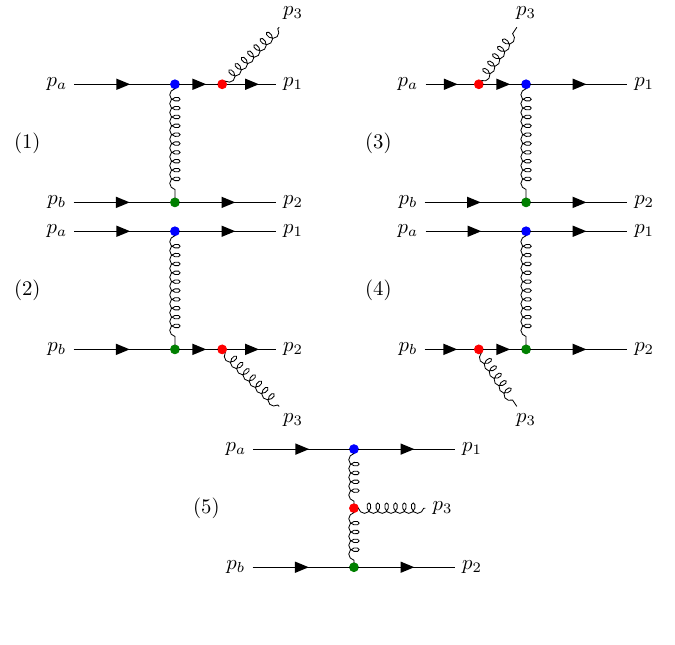}
    \hfill 
    \includegraphics[width=0.48\textwidth]{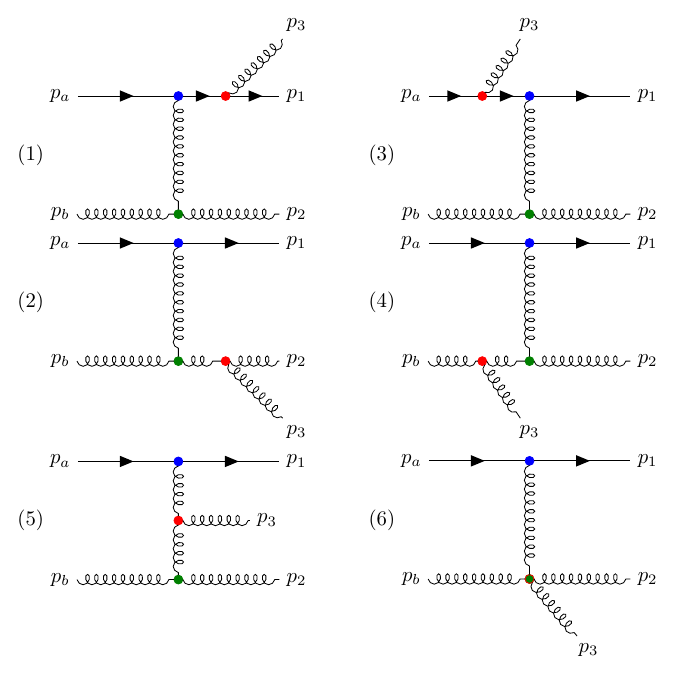}
    \caption{
        Feynman diagrams for the $t$ channel of the $q+q \to q+q+g$ (left) and $q+g \to q+g+g$ (right) processes. The green dots indicate the vertices corresponding to thermal partons. The blue dots indicate the vertices corresponding to the jet quark while the red dots denote the vertices corresponding to the emitted gluon. For the case of the 4-gluon point interaction (diagram 6 in the right figure) the medium parton vertex and the emitted gluon vertex are merged.
    }
    \label{fig:diags}
\end{figure*}

The corresponding Feynman diagrams for the $t$ channel of the $q+q \to q+q+g$ and $q+g \to q+g+g$ processes are illustrated in Fig. \ref{fig:diags}. Considering that the jet parton is a $u$-quark, the following inelastic reactions and channels are accounted for in this study:
\begin{itemize}
    \setlength\itemsep{-0.2em}
    \item $u+u \to u+u+g$ ($t + u$ channels)
    \item $u+\bar{u} \to u+\bar{u}+g$ ($t + s$ channels)
    \item $u+d \to u+d+g$ ($t$ channel)
    \item $u+\bar{d} \to u+\bar{d}+g$ ($t$ channel)
    \item $u+s \to u+s+g$ ($t$ channel)
    \item $u+\bar{s} \to u+\bar{s}+g$ ($t$ channel)
    \item $u+\bar{u} \to d+\bar{d}+g$ ($s$ channel)
    \item $u+\bar{u} \to s+\bar{s}+g$ ($s$ channel)
    \item $u+g \to u+g+g$ ($t + u + s$ channels).
\end{itemize}
We note that the $u$ channel for the $u+u \to u+u+g$ reaction gives practically identical contribution as the $t$ channel and is accounted for by doubling the $t$ channel contribution when evaluating the transport coefficient to simplify the calculations. The $s$ channels for all above reactions are strongly suppressed relative to $t$ channels and give only minor contributions to the transport coefficients. In case of the $u+g \to u+g+g$ reaction the $u$ channel is also strongly suppressed (cf. Ref. \cite{Grishmanovskii:2023gog}). Therefore, only the $t$ channels are considered in the current study. For the details of the evaluation of the $2\to 2$ transport coefficients and for the inelastic $2\to 3$ cross sections we refer the reader to Refs. \cite{Grishmanovskii:2022tpb, Grishmanovskii:2023gog}.

\subsection{Mass of the radiated gluon}

An important feature of the DQPM, contrary to the pQCD-based models, is the non-zero mass of the emitted gluon in the $2 \to 3$ reactions. In a thermal DQPM medium the mass of the radiated gluon in the on-shell case is assumed to be the gluon pole mass (obtained from Eq. \eqref{eq:Mg}). The jet parton, however, is not part of the QGP medium and should have a non-thermal mass. In turn, it is not clear whether the emitted gluon mass should still be thermal since the gluon can be emitted from a non-thermal jet parton. Therefore, in this work we also investigate the dependence of transport coefficients on the emitted gluon mass.

We note that the influence of the mass of the emitted gluon on the radiative energy loss of heavy quarks scattering with massless quarks and gluons from the QGP medium has been studied within scalar pQCD in Refs. \cite{Gossiaux:2010yx,Aichelin:2013mra,Gossiaux:2014jga}, where it has been shown that the emission rate decreases with increasing gluon mass.

Moreover, we indicate that all calculations below are presented for a quark jet with mass $M=0.01$ GeV. As shown in our previous study \cite{Grishmanovskii:2023gog}, at high collision energies the inelastic cross sections only slightly depend on the mass of jet parton.


\subsection{Effective coupling constants for jet elastic and inelastic scattering}
\label{sec:scenarios}

In this section, we consider five "scenarios" for different strong couplings for every possible vertex. As illustrated in Fig. \ref{fig:diags}, for every possible interaction channel for the $2 \to 3$ reaction in the $t$ channel of the $q+q \to q+q+g$ (left) and $q+g \to q+g+g$ (right) processes, one can define three different strong couplings: one associated with the thermal parton (green dot), one associated with the jet parton (blue dot), and one associated with the emitted gluon (red dot). We note that for the elastic $2 \to 2$ reactions the thermal (green) and jet (blue) vertices are the same in the corresponding Feynman diagrams as for the inelastic case.

Different scenarios (defined below) for the strong coupling $g$ are illustrated in Fig. \ref{fig:coupling_comparison} as a function of the temperature $T$ (upper plot) and jet energy $E$ (lower plot).


\subsubsection{Scenario 0: DQPM thermal \texorpdfstring{$g(T)$}{g(T)}}

This scenario corresponds to the default version of the DQPM where in all vertices the thermal DQPM coupling constant is taken, i.e.,
\begin{equation}
    g^{\text{DQPM}}(T) = g^{\text{DQPM}}(T,\mu_B=0),
\end{equation}
which is defined by Eq. (\ref{eq:coupling_DQPM}).

Since the DQPM coupling depends strongly on the medium temperature $T$ (cf. blue line in Fig. \ref{fig:coupling_comparison}) and grows rapidly in the vicinity of the critical temperature $T_c$, we expect a large value and strong temperature dependence of the transport coefficients for inelastic reactions similar to the elastic one, as shown in our previous study \cite{Grishmanovskii:2022tpb}. The DQPM coupling does not depend on the jet momentum or momentum transfer, so we expect a strong dependence for the inelastic transport coefficients on the initial jet energy $E$ similar to elastic case \cite{Grishmanovskii:2022tpb}.


\subsubsection{Scenario I: \texorpdfstring{$g = const$}{g=const}}

In this scenario, we investigate the transport coefficients without taking into account the direct influence of the coupling constant. For this we consider a constant value of the strong coupling $g$ for all vertices in $2\to 2$ and $2\to 3$ diagrams. The parton masses, however, are kept with the same temperature dependence as in the default DQPM.

The value of $g$ is chosen to match the commonly taken value $\alpha_s = 0.3$ (cf. the pQCD BAMPS model \cite{Fochler:2013epa}), i.e.,
\begin{equation}
    g = \sqrt{4\pi \cdot 0.3} \approx 1.94.
    \label{coupling_03}
\end{equation}
As one can see in Fig. \ref{fig:coupling_comparison}, although this value of $g$ (gray dotted lines) is smaller than the thermal $g^\mathrm{DQPM}(T)$ for temperatures up to $T \approx 0.6$ GeV, it is still larger than 1, so it must give a significant contribution to inelastic amplitudes, where it is accounted in each vertex. For this scenario, we expect a smaller value of $\qhat$ compared to Scenario 0, as well as a different form of the temperature dependence and energy dependence of transport coefficients.


\subsubsection{Scenario II: \texorpdfstring{$g(Q^2)$}{g(Q**2)} from the Zakharov model}

Motivated by Refs. \cite{Zakharov:2020whb, Zakharov:2020psr}, in this scenario we consider a momentum-dependent strong coupling frozen at low momenta at some value $\alpha_s^{fr}$:
\begin{equation}
    g^2(Q^2) = 
    \begin{cases}
        4\pi \alpha_s^{fr} 
        & \text{if } Q \le Q_{fr}, \\
        \frac{48\pi^2}{(11N_c-2N_f)\ln{\left (Q^2/\Lambda^2_{\text{QCD}}\right )}}
        & \text{if } Q > Q_{fr},
    \end{cases}
    \label{eq:coupling_Zakharov}
\end{equation}
with $\Lambda_{\text{QCD}}=0.2$ GeV, $Q_{fr} = \Lambda_{\text{QCD}} \exp(2\pi/9\alpha_s^{fr})$, and $\alpha_s^{fr}$ is a free parameter. In this work, we consider $\alpha_s^{fr} = 1.05$ (vacuum value) and $\alpha_s^{fr} = 0.42$ (in-medium value) taken from Ref. \cite{Zakharov:2020whb}. It is important to note that here thermal effects encompassed in $Q^2$ dependence suppress the in-medium QCD coupling.

For the jet vertex for both elastic and radiative collisions the $Q$ in Eq. \eqref{eq:coupling_Zakharov} is defined as the momentum transfer between the jet and the medium parton. The coupling for the thermal vertex remains to be $g^{\text{DQPM}}(T)$. For the emitted gluon in $2 \to 3$ reaction the value of $Q$ is suggested to be $k_T$ -- the transverse momentum of the emitted gluon.

Summing up, the choice of the couplings for the inelastic processes for Scenario II is the following:
\begin{itemize}
    \item [\textcolor{Green}{$\bullet$}] thermal vertex -- $g^\mathrm{DQPM} (T)$.
    \item [\textcolor{blue}{$\bullet$}] jet vertex -- $g(Q^2)$ with $Q$ being the momentum transfer between the jet and the medium parton.
    \item [\textcolor{red}{$\bullet$}] radiative vertex -- $g(k_T^2)$, where $k_T$ is the transverse momentum of the emitted gluon.
\end{itemize}


\subsubsection{Scenario III: QLBT model}

Another way to define strong couplings has been introduced in the QLBT model \cite{Liu:2021dpm, Liu:2023rfi}, where the couplings associated with the jet parton and the emitted gluon vertices are assumed to have the following parametric form:
\begin{equation}
    g^2(E) = \frac{48 \pi^2}{(11N_c-2N_f) \ln\left[(AE/T_c+B)^2\right]},
    \label{eq:coupling_QLBT}
\end{equation}
where $E$ is the jet parton energy in the rest frame, $T_c = 150$ MeV is the critical temperature, and the parameters $A,B$ are determined from the heavy quark observables (such as $R_{AA}$ and $v_2$). The coupling for the thermal vertex remains to be $g^{\text{DQPM}}(T)$. The coupling $g^2(E)$ is shown in Fig. \ref{fig:coupling_comparison} by the green lines: dashed-dotted for $E=10$ GeV and dashed for $E=100$ GeV on the upper plot, showing the temperature dependence, and by the dash-dotted line in the lower plot, showing the jet energy dependence.

The choice of the couplings for Scenario III is the following:
\begin{itemize}
    \item [\textcolor{Green}{$\bullet$}] thermal vertex -- $g^\mathrm{DQPM} (T)$.
    \item [\textcolor{blue}{$\bullet$} \textcolor{red}{$\bullet$}] jet/radiative vertices -- $g(E)$, $E$ is the jet energy.
\end{itemize}


\subsubsection{Scenario IV: DREENA framework}

The last scenario is motivated by the couplings used in the DREENA framework \cite{Zigic:2021rku,Karmakar:2023ity}, where the jet interaction with the thermal QCD medium is computed within the two-loop HTL model \cite{DJORDJEVIC2014286, Djordjevic:2009cr} and implies the following form of the coupling:
\begin{equation}
    g^2(t)=\frac{48\pi^2}{(11N_c-2N_f)}\frac{1}{\ln{(\frac{t}{\Lambda^2})}},
    \label{eq:coupling_DREENA}
\end{equation}
with $\Lambda = 0.2$ GeV.

The choice of the couplings for Scenario IV is the following:
\begin{itemize}
    \item [\textcolor{Green}{$\bullet$}] thermal vertex -- $g^\mathrm{DQPM} (T)$.
    \item [\textcolor{blue}{$\bullet$}] jet vertex -- $g(ET)$ according to Eq.(\ref{eq:coupling_DREENA}) with $t=ET$, where $E$ is the jet energy.
    \item [\textcolor{red}{$\bullet$}] radiative vertex -- $g(Q^2)$ according to Eq.(\ref{eq:coupling_DREENA}) with $t=Q^2$, where $Q$ is the virtuality of the intermediate parton before(after) the gluon emission.
\end{itemize}

\begin{figure}[ht!]
    \centering
    \includegraphics[width=\columnwidth]{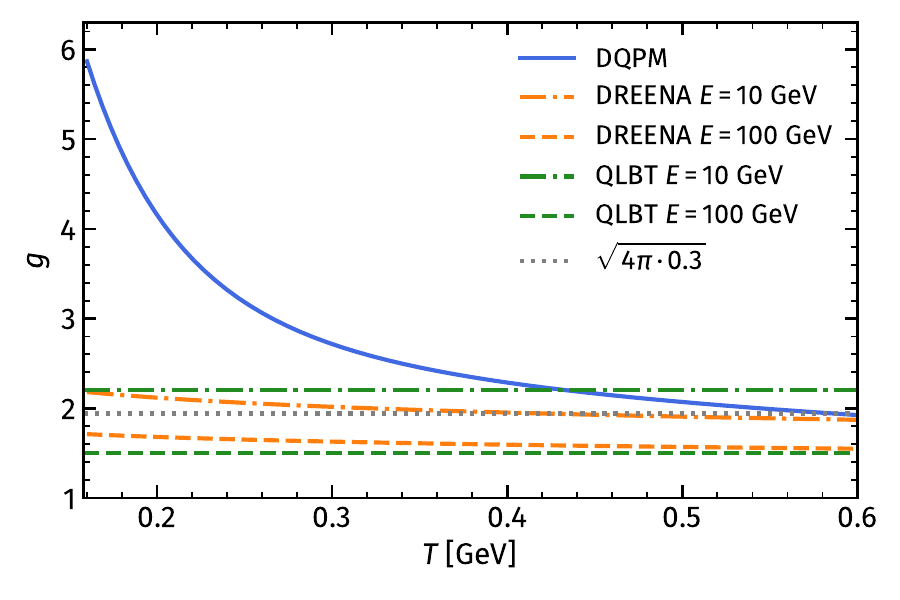}
    \includegraphics[width=\columnwidth]{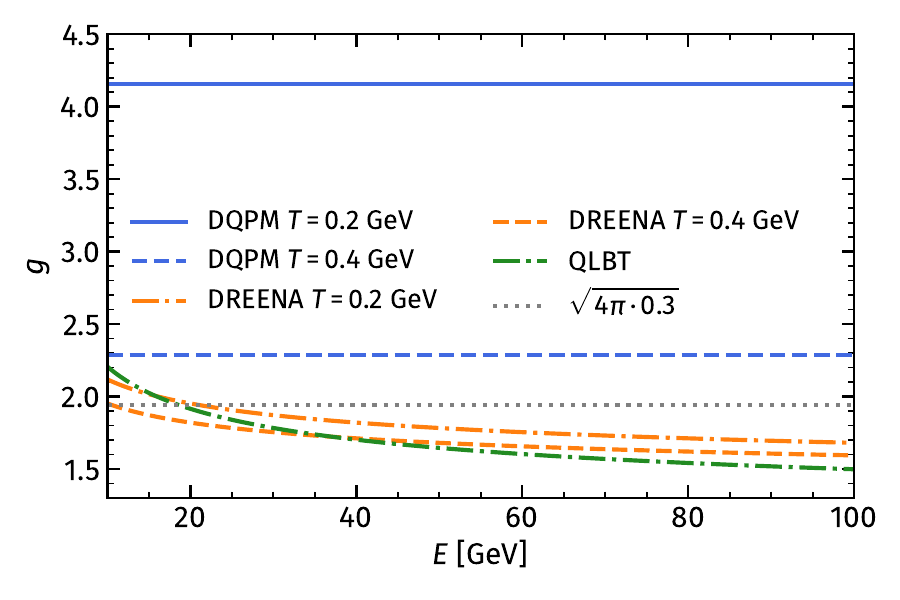}
    \caption{
        Strong coupling $g$ as a function of the temperature $T$ (upper) and jet energy $E$ (lower). Blue lines correspond to the DQPM coupling defined by Eq. \eqref{eq:coupling_DQPM}, orange lines show the DREENA coupling defined by Eq. \eqref{eq:coupling_DREENA}, green lines display the QLBT coupling defined by Eq. \eqref{eq:coupling_QLBT}, and the gray dotted line shows $g = \sqrt{4\pi \cdot 0.3}$.
    }
    \label{fig:coupling_comparison}
\end{figure}

The orange lines in Fig. \ref{fig:coupling_comparison} show the coupling $g^2(E)$: dot-dashed for $E=10$ GeV and dashed for $E=100$ GeV on the upped plot for the $T$ dependence, and dot-dashed for $T=0.2$ GeV and dashed for $T=0.4$ GeV in the lower plot for the $E$ dependence.


\subsection*{Summary table for Scenarios 0 - IV}

An overview of the presented scenarios is shown in Table \ref{tbl:scenarios-alphas}.

\begin{table}[ht!]
    \centering
    \scalebox{0.94}{
    \begin{tabular}{|l|c|c|c|}
        \hline
        & \multicolumn{3}{c|}{\textbf{Vertex}}
        \\ \hline
        \textbf{Model} & \textcolor{Green}{$\bullet$} medium parton\; & \textcolor{blue}{$\bullet$} jet parton\; & \textcolor{red}{$\bullet$} emitted gluon\;
        \\ \hline \hline
        Scenario 0 & \multicolumn{3}{c|}{$g^{\text{DQPM}}(T)$}
        \\ \hline
        Scenario I & \multicolumn{3}{c|}{$g = \sqrt{4 \pi \cdot 0.3}$}
        \\ \hline
        Scenario II & $g^{\text{DQPM}}(T)$ & $g(Q^2)$, Eq. \eqref{eq:coupling_Zakharov} & $g(k_T^2)$, Eq. \eqref{eq:coupling_Zakharov} 
        \\ \hline
        Scenario III & $g^{\text{DQPM}}(T)$ & $g(E)$, Eq. \eqref{eq:coupling_QLBT} & $g(E)$, Eq. \eqref{eq:coupling_QLBT} 
        \\ \hline
        Scenario IV & $g^{\text{DQPM}}(T)$ & $g(ET)$, Eq. \eqref{eq:coupling_DREENA} & $g(Q^2)$, Eq. \eqref{eq:coupling_DREENA} 
        \\ \hline
    \end{tabular}
    }
    \caption{Scenarios for the effective coupling.}
    \label{tbl:scenarios-alphas}
\end{table}

\section{Results: radiative differential cross sections}
\label{sec:res_dxs}

\begin{figure*}[ht!]
    \centering
    \includegraphics[width=0.9\columnwidth]{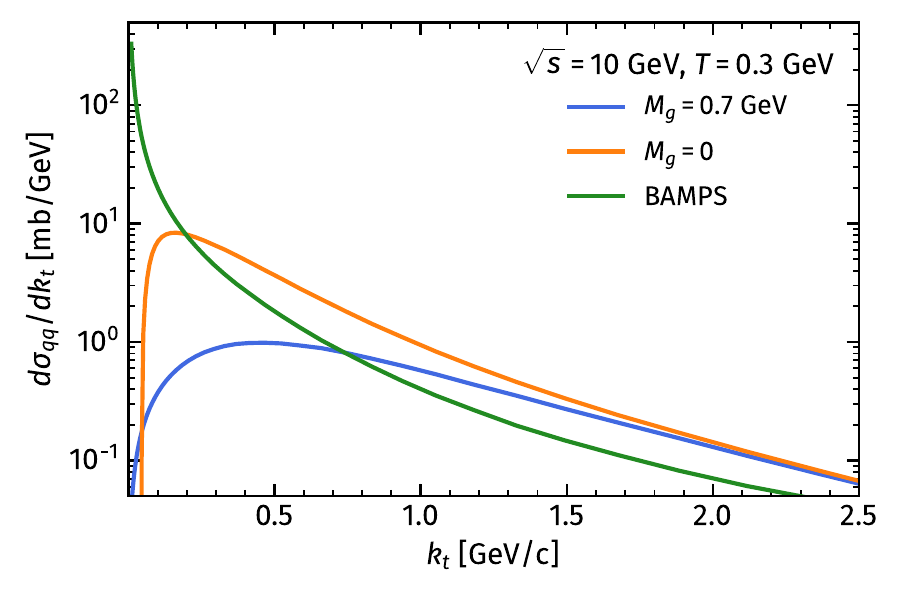}
    \includegraphics[width=0.9\columnwidth]{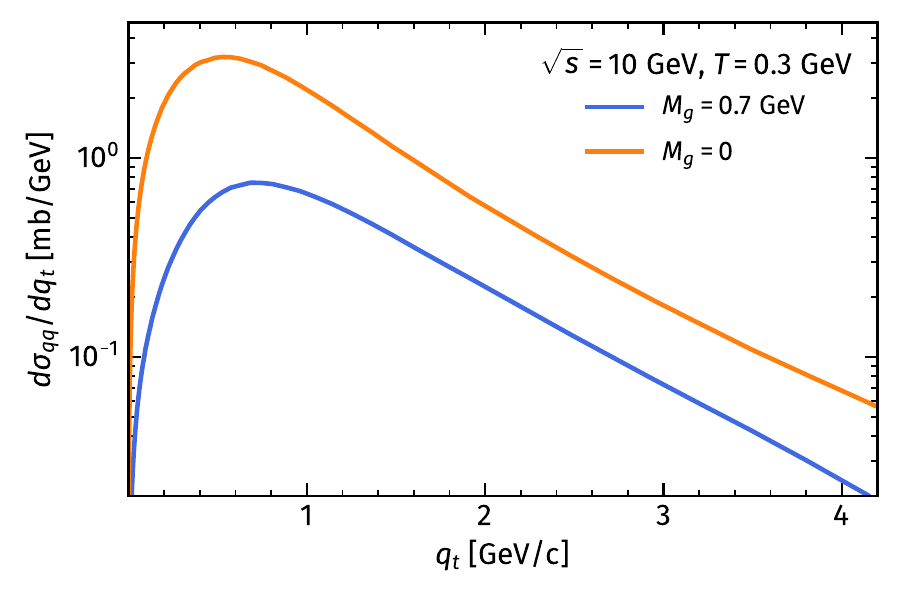}
    \includegraphics[width=0.9\columnwidth]{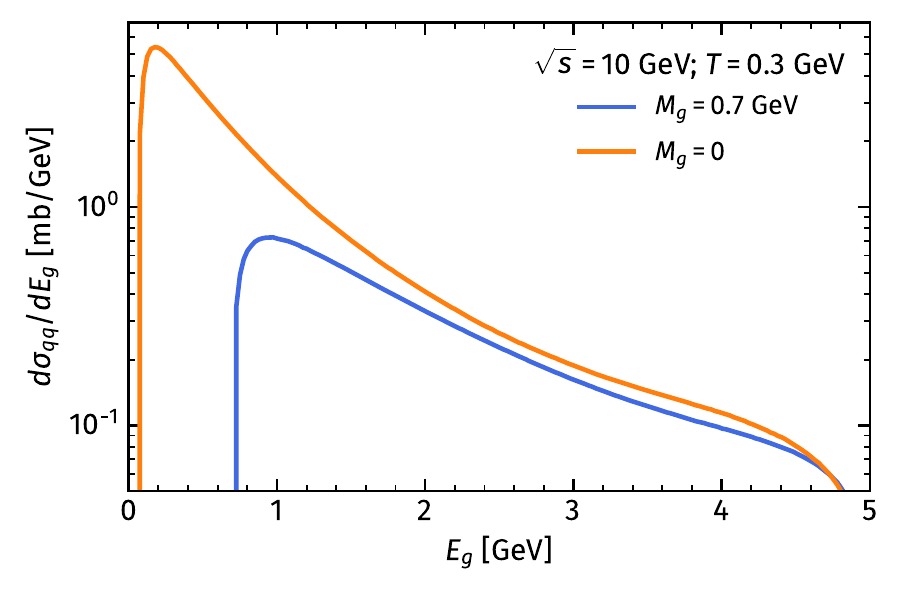}
    \includegraphics[width=0.9\columnwidth]{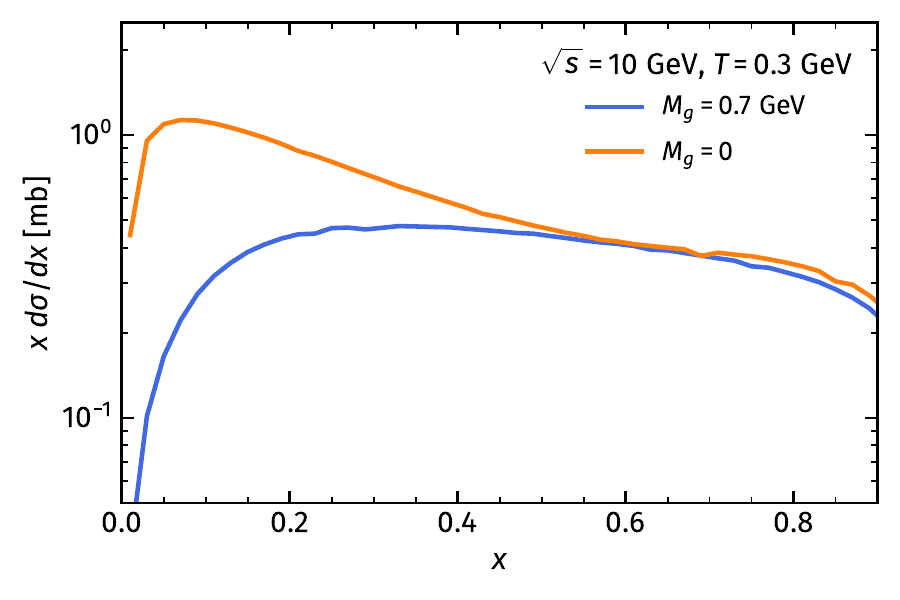}
    \caption{
        DQPM differential cross sections: $d\sigma/dk_t$ as a function of the transverse momentum $k_t$ of the emitted gluon (upper left), $d\sigma/dq_t$ as a function of the transverse momentum $q_t$ of the jet (upper right), $d\sigma/dE_g$ as a function of the energy $E_g$ of the emitted gluon (lower left), and $x\ d\sigma/dx$ as a function of the longitudinal ratio $x = p_L^{g}/p_L^{j}$ (lower right) for $q+q \to q+q+g$ scattering at $\sqrt{s}=10$ GeV and $T=0.3$ GeV in case of massive (thermal) emitted gluon (blue lines) and massless emitted gluon (orange lines). The green line on the upper left plot shows $d\sigma/dk_t$ calculated with the pQCD BAMPS model for $\alpha_s=0.3$ and massless partons without LPM cutoff.
    }
    \label{fig:dXSi}
\end{figure*}

We recall that the $2\to 3$ cross sections and interaction rates for $q+q \to q+q+g$ and $q+g\to q+g+g$ scattering have been calculated within the DQPM in Ref. \cite{Grishmanovskii:2023gog} and compared to the elastic ones. It has been shown that the inelastic reactions contribute only little to the interaction rate $\Gamma$ of the QGP in thermal equilibrium. That is due to the fact that partonic scatterings at larger $\sqrt{s}$ are strongly suppressed in the thermal QGP -- cf. Fig. 17 of Ref. \cite{Grishmanovskii:2023gog}, thus, only low-$\sqrt{s}$ elastic and inelastic reactions dominantly contribute to the interaction rates, where the elastic cross sections are much larger than the inelastic ones. However, inelastic contribution is expected to be important for the calculation of the energy loss of the off-equilibrium fast jet parton propagating through the thermal QGP medium, since the inelastic $q+q$ and $q+g$ cross sections grow with increasing collision energy \cite{Grishmanovskii:2023gog}.

As mentioned above, while the DQPM considers the properties of all colliding partons to be in thermal equilibrium, a fast equilibration of the emitted gluon after the first inelastic collision is questionable. Thus, it is interesting to study how the differential cross sections -- related to the energy loss of the jet parton -- and the momentum distribution of the emitted gluon are sensitive to the mass of the radiated gluon.

In Fig. \ref{fig:dXSi} we present the different types of differential DQPM cross sections for the case of thermal massive -- with the pole DQPM mass $M_g(T)$ defined by Eq. (\ref{eq:Mg}) -- (blue lines) and massless (orange lines) emitted gluon for the $q+q \to q+q+g$ scattering at $\sqrt{s}=10$ GeV and $T=0.3$ GeV. The upper left plot shows the differential cross sections $d\sigma/dk_t$ as a function of the transverse momentum $k_t$ of the emitted gluon. The upper right plot indicates the differential cross section $d\sigma/dq_t$ as a function of the transverse momentum $q_t$ of the jet. The lower left plot displays the differential cross section $d\sigma/dE_g$ as a function of the energy $E_g$ of the emitted gluon. The lower right plot demonstrates the distribution $x\ d\sigma/dx$ as a function of the longitudinal ratio $x = p_L^{g}/p_L^{j}$, where $p_L^{g}$ and $p_L^{j}$ are the longitudinal momenta of gluon and initial jet parton, respectively.

One can see from Fig. \ref{fig:dXSi}, all DQPM differential cross sections for radiation of massless gluons at small $k_t, q_t, E_g$, or $x$ are larger than for the massive cases. This is due to the fact that the kinematically available phase space is strongly reduced for the emission of massive heavy gluons compared to the massless ones. Therefore, it is expected to obtain a similar increase of transport coefficients for massless emitted gluons versus massive ones. Our results are consistent with the findings in Refs. \cite{Gossiaux:2010yx, Aichelin:2013mra, Gossiaux:2014jga} for the radiative energy loss of heavy quarks in the QGP medium described by scalar pQCD, where it has been shown that the gluon emission cross section decreases with increasing gluon mass.

In the upper left plot we compare the DQPM $d\sigma/dk_t$ to the pQCD differential cross section within the BAMPS framework with $\alpha_s=0.3$ and massless partons (green line), which diverges at $k_t \to 0$ without LPM effect \cite{Senzel:2020tgf}. Thus, the pQCD $d\sigma/dk_t$ is much larger at low $k_t$ and decreases much faster with increasing $k_t$ relative to the DQPM cross sections.


\section{Results: \texorpdfstring{$\qhat$}{qhat} and energy loss \texorpdfstring{$dE/dx$}{dE/dx}}
\label{sec:res_qhat}

Here we present the $T$ and $p$ dependence of the jet transport coefficients -- $\qhat$ and energy loss $dE/dx$ -- for elastic and inelastic reactions for different scenarios for the strong coupling as presented in Sec. \ref{sec:scenarios} and compare our results with previous estimations from the literature. Also we study the dependence of transport coefficients on the mass of emitted gluons within the DQPM.

We will show the results for the elastic and inelastic (or radiative) $\qhat$ coefficient as well as for the sum of both elastic and inelastic contributions:
$$\qhat=\qhat^\text{elastic} + \qhat^\text{inelastic};$$ 
similar holds for the energy loss:
$$dE/dx = (dE/dx)^\text{elastic} + (dE/dx)^\text{inelastic}.$$


\subsection{Dependence of transport coefficients \texorpdfstring{$\qhat$}{qhat} and \texorpdfstring{$dE/dx$}{dE/dx} on the mass of the emitted gluon}

\begin{figure*}[!t]
    \centering
    \includegraphics[width=\columnwidth]{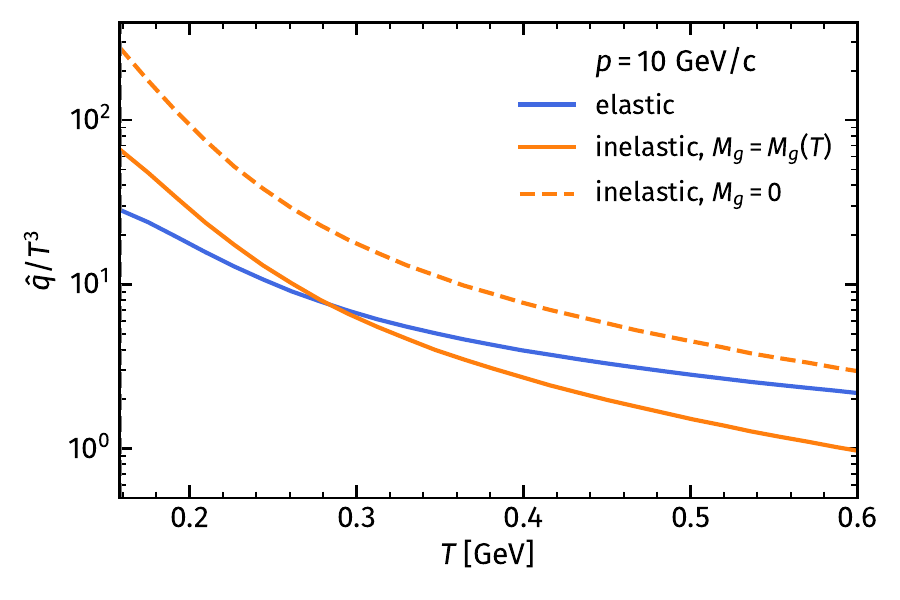}
    \includegraphics[width=\columnwidth]{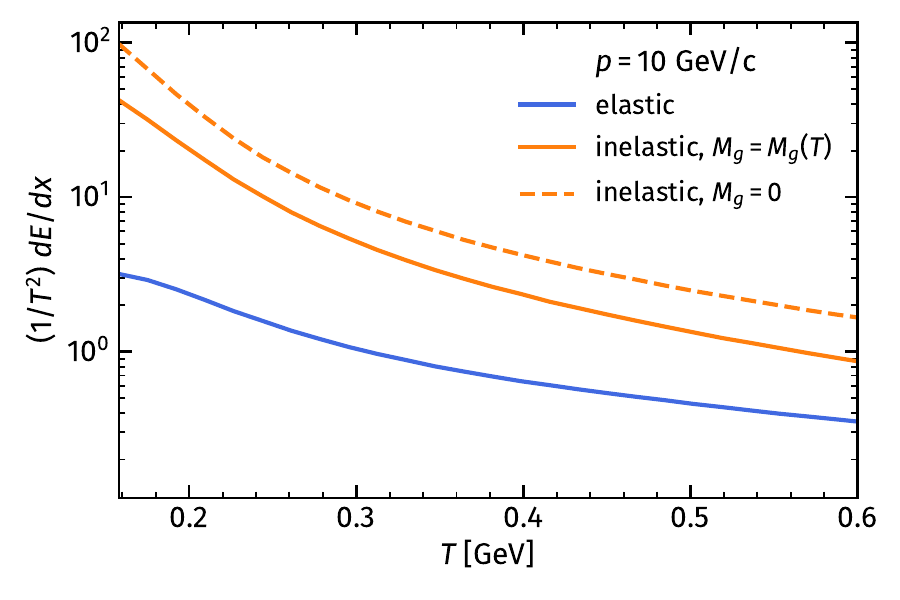}
    \includegraphics[width=\columnwidth]{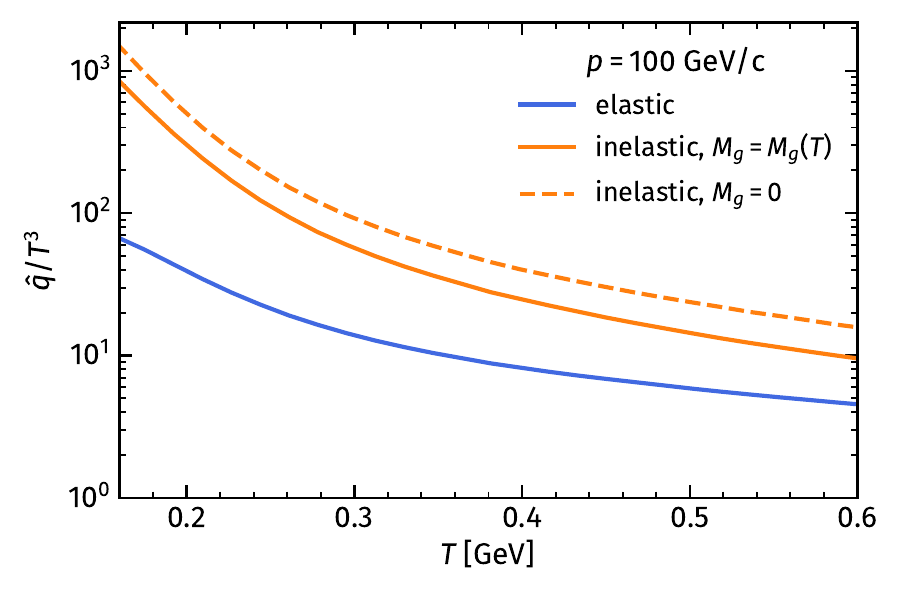}
    \includegraphics[width=\columnwidth]{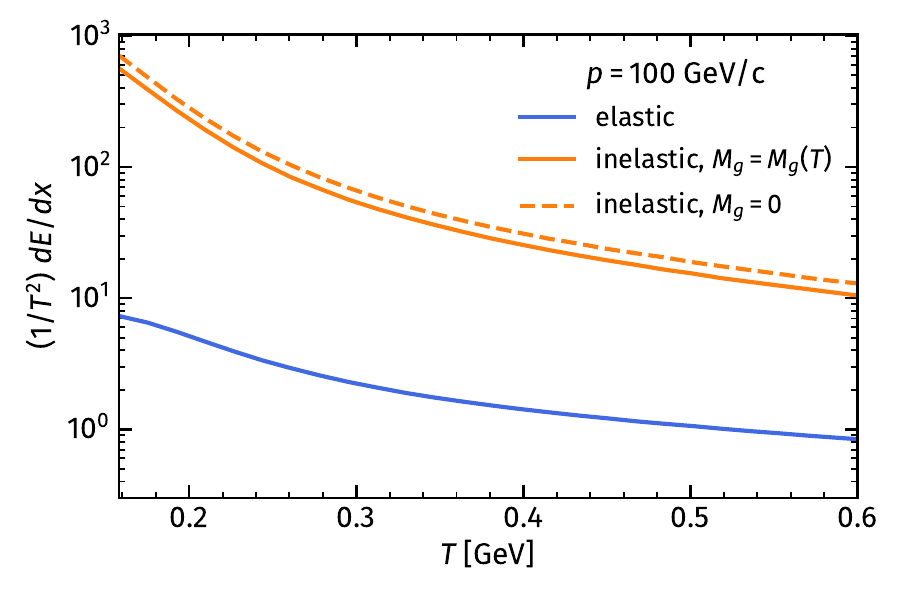}
    \caption{
        Temperature dependence of $\qhat/T^3$ (left plots) and the energy loss $(1/T^2) dE/dx$ (right plots) for elastic (blue lines) and inelastic (orange lines) scattering for different momenta of the jet parton: $p= 10 $ GeV$/c$ (upper plots) and 100 GeV$/c$ (lower plots). The inelastic transport coefficients for the emission of a massive gluon are shown by solid lines, while for a massless ($M_g=0$) gluon by dashed lines.
    }
    \label{fig:qhat-T_el-vs-in}
    \centering
    \includegraphics[width=\columnwidth]{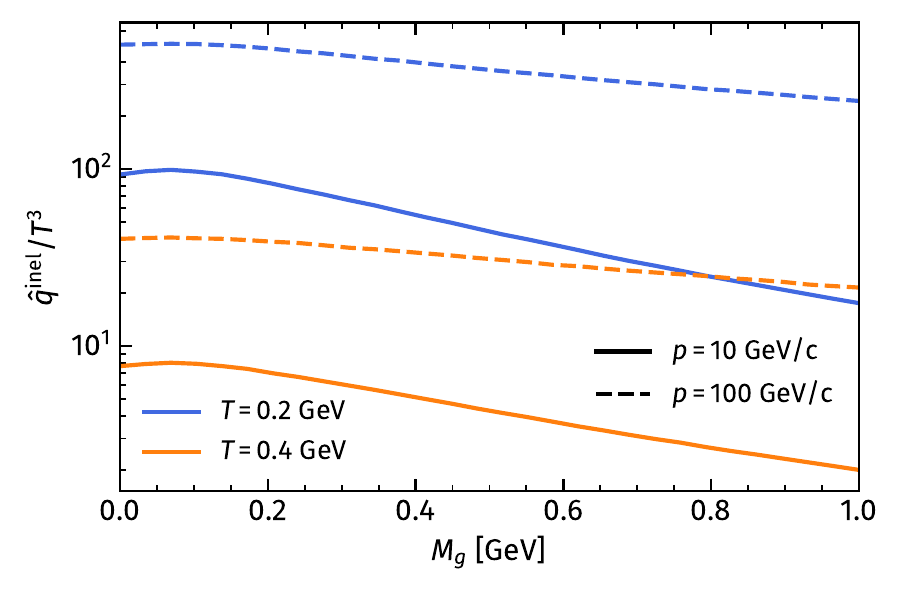}
    \includegraphics[width=\columnwidth]{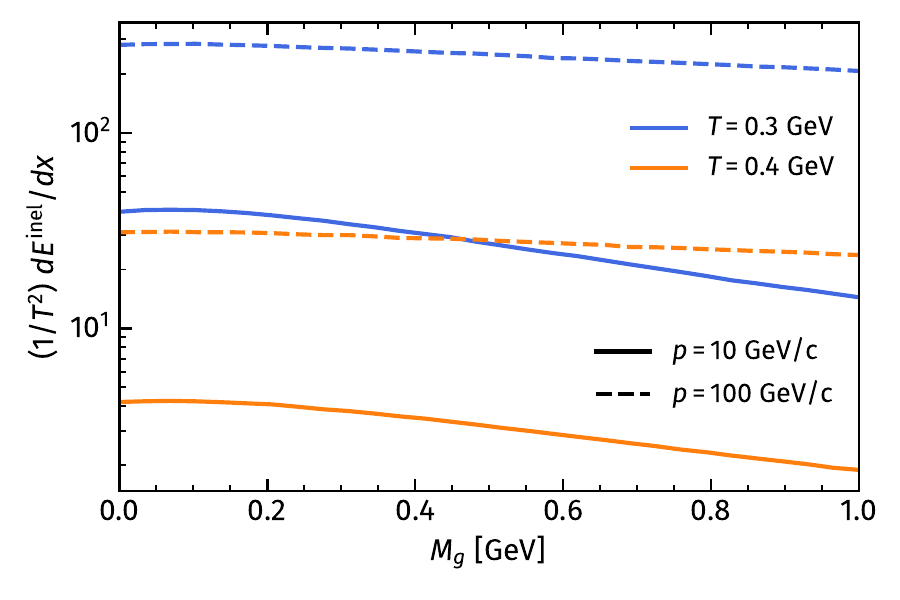}
    \caption{
        Scaled inelastic jet transport coefficient $\qhat/T^3$ (left plot) and the energy loss $(1/T^2) dE/dx$ (right plot) as functions of the emitted gluon mass $M_g$ for different momenta of jet $p= 10$ GeV$/c$ (solid lines) and $100$ GeV$/c$ (dashed lines) for two temperatures $T=0.2$ GeV (blue lines) and $T=0.4$ GeV (orange lines).
    }
    \label{fig:qhat-Mg}
\end{figure*}

In this section we study the influence on the mass of emitted gluons from inelastic scattering of the jet parton with the thermal sQGP medium -- described by the DQPM quasiparticles -- on the transport coefficients $\qhat$ and energy loss $dE/dx$. For this study we consider the pure DQPM model (Scenario 0).

Figure \ref{fig:qhat-T_el-vs-in} shows the temperature dependence of the scaled $\qhat$ coefficient (left plots) and the scaled energy loss $dE/dx$ (right plots) for elastic (blue lines), inelastic scattering (orange lines) with radiation of a massive gluon (solid lines) and massless ($M_g=0$) gluon (dashed lines) for two quark jet momenta: $p=10$ GeV$/c$ (upper plots) and $p=100$ GeV$/c$ (lower plots). Similar to differential cross sections, the $\qhat$ and $dE/dx$ increase as the gluon mass decreases. Moreover, for the massless emitted gluon the inelastic $\qhat$ and $dE/dx$ become significant even for small jet momenta of 10 GeV$/c$.

As follows from Fig. \ref{fig:qhat-T_el-vs-in}, the inelastic reactions contribute dominantly to $\qhat$ and $dE/dx$. As mentioned above, in a thermalized QGP the contribution of inelastic reactions to the interaction rate is small due to predominantly low $\sqrt{s}$ collisions of thermal partons. However, this changes for the scattering of a fast jet quark with a thermal parton at large $\sqrt{s}$. Since the inelastic $q+q$ and $q+g$ cross sections grow faster with increasing $\sqrt{s}$ than the elastic ones, the inelastic energy loss and transverse momentum broadening dominate the elastic ones for high momentum jet partons propagating through the thermalized sQGP. The available collision energy is sufficient for the emission of massless pQCD gluons (dashed lines) as well as for the heavy DQPM gluons (solid lines).

Figure \ref{fig:qhat-Mg} displays the inelastic $\qhat/T^3$ coefficient (left plot) and energy loss $(1/T)^2 dE/dx$ (right plot) as functions of the emitted gluon mass $M_g$. For all displayed temperatures and jet momenta the values of $\qhat$ and energy loss $dE/dx$ monotonically decrease with increasing of the emitted gluon mass.


\subsection{Comparison of jet transport coefficients with different scenarios for the effective coupling constant}

Here we investigate the influence of the effective coupling $g$ in thermal, jet, and radiative vertices on the temperature and energy dependence of $\qhat$ and $dE/dx$ within the different Scenarios 0-IV as described in Sec. \ref{sec:scenarios}. We stress that all calculations of Feynman diagrams are presented within the "standard" DQPM propagators, and masses of quasiparticles are taken at the pole mass of the spectral function, i.e., only the coupling constants in the vertices are varied -- cf. Table \ref{tbl:scenarios-alphas} and Fig. \ref{fig:coupling_comparison}. 

\begin{figure*}[ht!]
    \centering
    \includegraphics[width=0.8\textwidth]{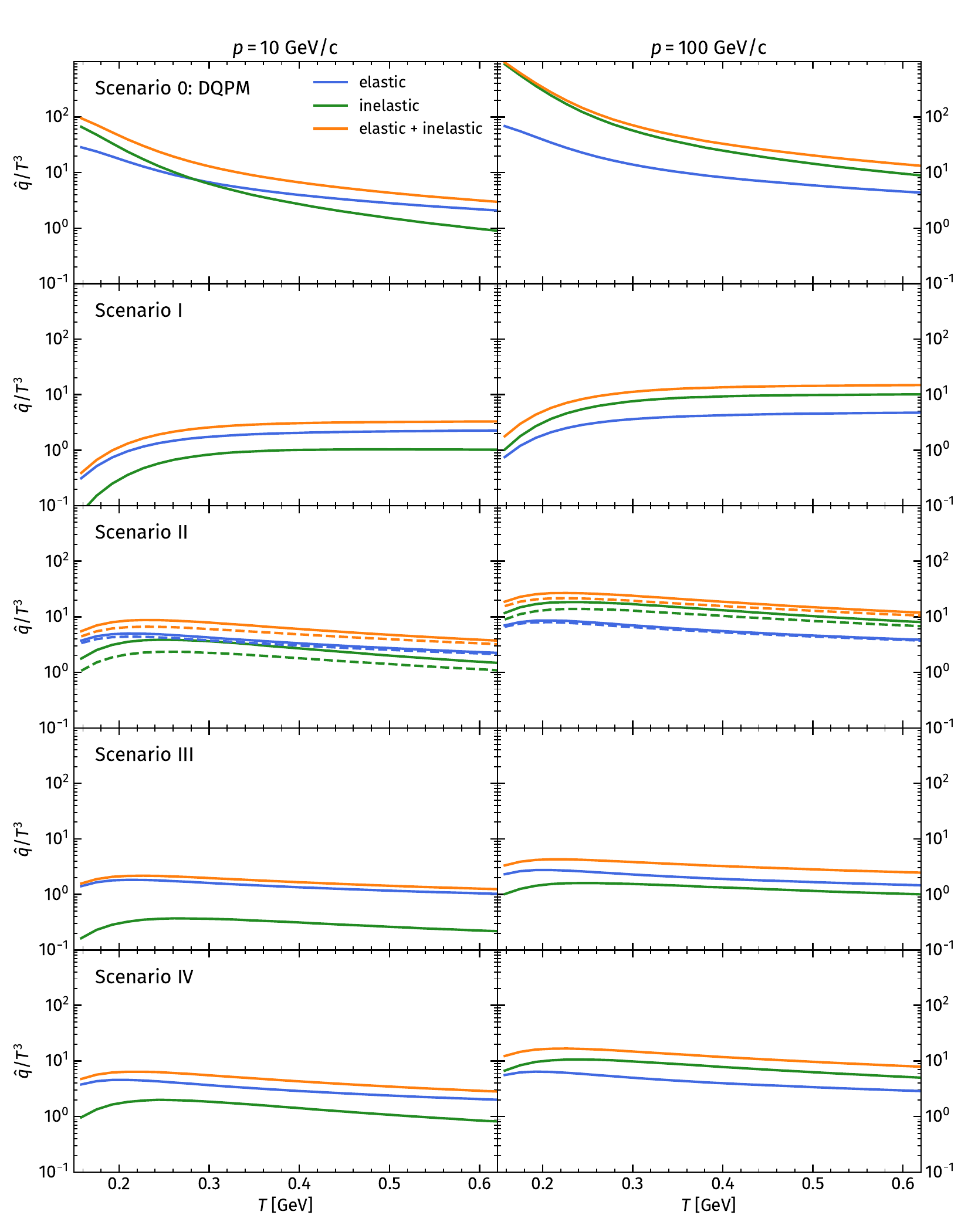}
    \caption{
        Temperature dependence of the scaled $\qhat$ coefficient for Scenarios 0-IV (from top to bottom) for the strong coupling for two momenta of the jet parton of $p=10$ GeV$/c$ (left plots) and $p=100$ GeV$/c$ (right plots). Blue lines represent the results for elastic processes only, while orange lines represent the results for the sum of elastic and inelastic contributions.
    }
    \label{fig:qhat-T_scenarios}
\end{figure*}

\begin{figure*}[ht!]
    \centering
    \includegraphics[width=0.8\textwidth]{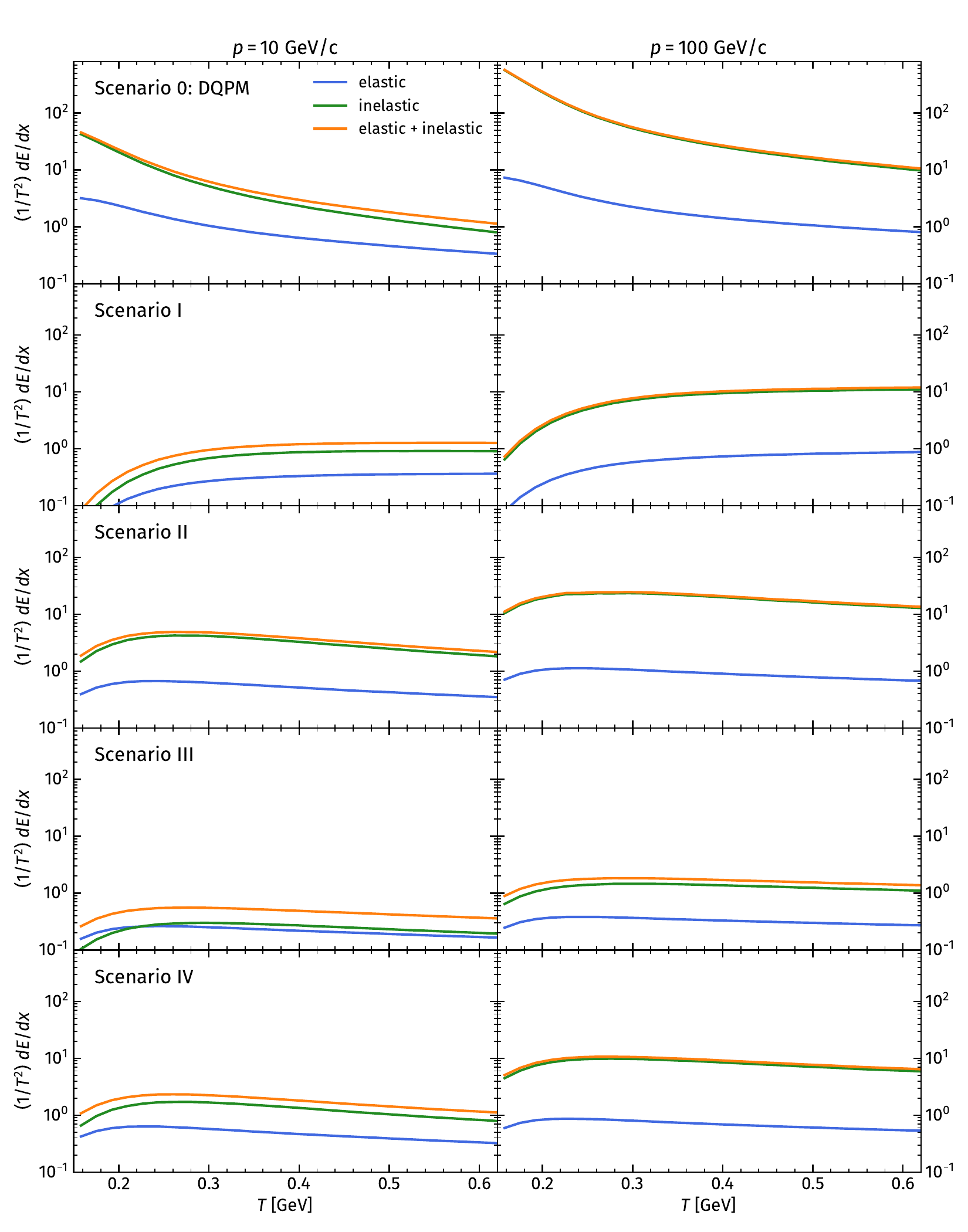}
    \caption{
        Temperature dependence of the scaled energy loss $dE/dx$ for different Scenarios 0-IV for two momenta of the jet parton with $p=10$ GeV$/c$ (left side) and $p=100$ GeV$/c$ (right side). Blue lines represent results for elastic processes only, while orange lines represent results for the sum of elastic and inelastic contributions.
    }
    \label{fig:elos-T_scenarios}
\end{figure*}

In Figs. \ref{fig:qhat-T_scenarios} and \ref{fig:elos-T_scenarios} we show the temperature dependence of the scaled $\qhat$ and $dE/dx$, respectively, for Scenarios 0-IV (from top to bottom) for two momenta of the jet parton of $p=10$ GeV$/c$ (left plots) and $p=100$ GeV$/c$ (right plots). The blue lines represent the scaled transport coefficients for elastic processes only, while the orange lines stand for the results of the sum of elastic and inelastic contributions.

$\bullet$ The upper rows in Figs. \ref{fig:qhat-T_scenarios} and \ref{fig:elos-T_scenarios} show the standard DQPM results (Scenario 0), where for all vertices the thermal $g(T)$ is used. As seen from these plots, accounting for the inelastic scattering substantially increases $\qhat/T^3$ and $(1/T^2)dE/dx$, especially at low temperatures due to the strong $T$ dependence of the strong coupling $g(T)$ -- cf. Fig. \ref{fig:coupling_comparison}.

$\bullet$ The second row in Figs. \ref{fig:qhat-T_scenarios} and \ref{fig:elos-T_scenarios} show the results for the Scenario I, where all vertices (thermal, jet parton, emitted gluon) have been taken as a constant $g=\sqrt{4\pi \cdot 0.3}$, i.e., $\alpha_s=0.3$. This scenario allows to illustrate explicitly the influence of the choice of the coupling $g$ on $\qhat$ and $dE/dx$. One can see that the scaled $\qhat$ decreases with decreasing $T$, so the strong rise of $\qhat$ for Scenario 0 at low $T$ comes from the increase of the DQPM strong coupling $g(T)$ in the vicinity of $T_c$, as illustrated in Fig. \ref{fig:coupling_comparison}. The same holds for the energy loss $dE/dx$, however, the relative contribution of inelastic reactions is larger than that for $\qhat$, and the $p$ dependence is stronger.

$\bullet$ The third row in Figs. \ref{fig:qhat-T_scenarios} and \ref{fig:elos-T_scenarios} shows the Scenario II, where the coupling constants in the jet parton and emitted gluon vertices are taken from the Zakharov model \cite{Zakharov:2020whb, Zakharov:2020psr} and depend on the squared momenta in the corresponding vertex -- $g(Q^2)$ or $g(k_T^2)$. The scaled $\qhat$ for this scenario shows a small rise with decreasing $T$. The dashed lines show the dependence of $\qhat$ on the choice of the parameter $\alpha_s^{fr}$ in Eq. (\ref{eq:coupling_Zakharov}): the solid lines stand for $\alpha_s^{fr} = 1.05$ (vacuum value), while the dashed line indicates the results for $\alpha_s^{fr} = 0.42$ (in-medium value). One can see that the in-medium modification of $\alpha_s^{fr}$ only slightly reduces $\qhat$. This reduction is more prominent for inelastic reactions depicted by the green lines. Again, similar observations are valid for $dE/dx$ with stronger $T$ and $p$ dependencies.

$\bullet$ The fourth row in Figs. \ref{fig:qhat-T_scenarios} and \ref{fig:elos-T_scenarios} represent the Scenario III with an energy dependent $g(E)$ -- in line with the QLBT model \cite{Liu:2021dpm, Liu:2023rfi} -- in the jet and radiative vertices. This leads to a quite flat $\qhat$ of smaller value as well as only a small increase of $\qhat$ with jet energy (cf. left and right plots) compared to the other scenarios due to the fact the $g(E)$ is smaller than $g$ from the other scenarios; it does not depend on $T$ and decrease with $E$ as demonstrated in Fig. \ref{fig:coupling_comparison} by the green lines. For $dE/dx$ we observe a similar trend, however, the relative contribution of inelastic reactions is smaller than for the other scenarios.

$\bullet$ The fifth row in Figs. \ref{fig:qhat-T_scenarios} and \ref{fig:elos-T_scenarios} displays the Scenario IV with the couplings used in the DREENA model \cite{Zigic:2021rku, Karmakar:2023ity, DJORDJEVIC2014286, Djordjevic:2009cr}, implying the HTL interaction of the jet quark with the thermal QCD medium. Here the jet vertex is replaced by the $g(ET)$ coupling, and the radiative vertex by $g(Q^2)$, where $Q^2$ stands for the virtuality of the intermediate parton before(after) the gluon emission. In this scenario the behavior of $\qhat$ and $dE/dx$ is similar to Scenario III and dominated by the energy dependence of the coupling $g(ET)$ -- cf. orange lines in Fig. \ref{fig:coupling_comparison}.

\begin{figure}[ht!]
    \centering
    \includegraphics[width=\columnwidth]{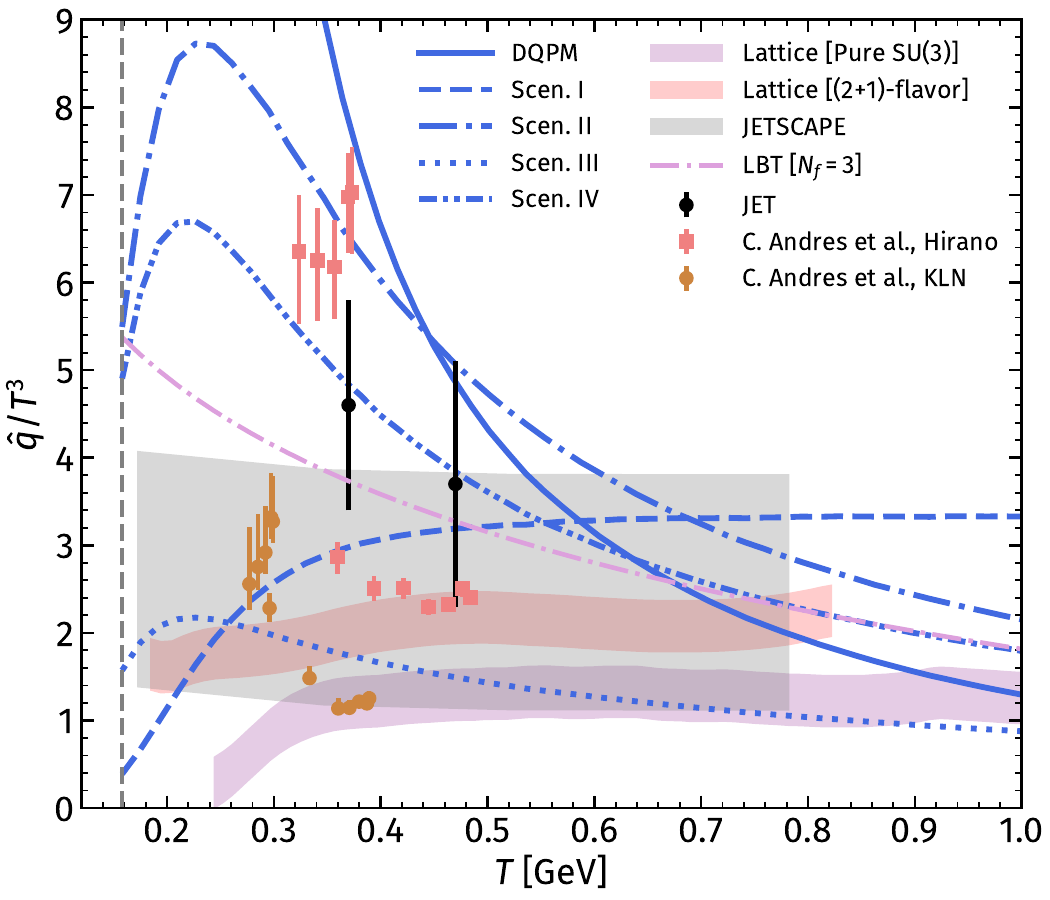}
    \caption{
        Temperature dependence of the scaled jet transport coefficient $\qhat/T^3$. Blue lines describe the DQPM results for a quark jet with mass $M=0.01$ GeV and energy $E=10$ GeV for the scenarios described in Sec. \ref{sec:scenarios}. The pink dash-dotted line represents the LBT results for $N_f=3$ and $p=10$ GeV$/c$ \cite{He:2015pra}, while the red (upper) and purple (lower) areas show lQCD estimates \cite{kumar2020jet} for pure SU(3) gauge theory and (2+1) flavor QCD, respectively, in the limit of an infinitely hard jet parton. The gray area corresponds to the results from the JETSCAPE Collaboration ($p = 100$ GeV$/c$) \cite{JETSCAPE:2021ehl}. The black dots represent the phenomenological extraction by the JET Collaboration presented for $p=10$ GeV$/c$ \cite{Burke:2013yra}. The yellow and red dots display results from \cite{Andres:2016iys}. The vertical gray dashed line indicates the critical temperature $T_c=0.158$ GeV.
    }
    \label{fig:qhat-T_cummulative}
\end{figure}

The results for the temperature dependence of the elastic + inelastic $\qhat/T^3$ for Scenarios 0-IV (for the momentum of a jet parton of $p=10$ GeV$/c$) are combined now in Fig. \ref{fig:qhat-T_cummulative} in terms of blue lines (cf. legend) and compared to the different models: the pink dash-dotted line represents the LBT results for $N_f=3$ and $p=10$ GeV$/c$ \cite{He:2015pra}, while the red (upper) and purple (lower) areas represent lQCD estimates \cite{kumar2020jet} for pure SU(3) gauge theory and (2+1) flavour QCD, respectively, in the limit of an infinitely hard jet parton. The gray area corresponds to the results from the JETSCAPE Collaboration ($p = 100$ GeV$/c$) \cite{JETSCAPE:2021ehl}. The black dots show the phenomenological extraction by the JET Collaboration presented for $p=10$ GeV$/c$ \cite{Burke:2013yra}. The yellow and red dots represent results from the phenomenological extraction within the BDMPS-Z quenching formalism using data of inclusive particle suppression at RHIC and LHC energies for the two distinct hydro frameworks from Ref. \cite{Andres:2016iys} for dynamical evolution -- "Hirano" and KLN. The vertical gray dashed line indicates the critical temperature $T_c=0.158$ GeV.

As follows from Fig. \ref{fig:qhat-T_cummulative}, the behavior of the scaled $\qhat$ is dominated by the choice of the strong coupling $g$ and deviates very strongly, especially at low temperatures $T$. While Scenario I with a fixed $\alpha_s=0.3$ shows a decrease of $\qhat$ at low $T$, the DQPM (Scenario 0) shows a strong rise due to the thermal $g(T)$ for all vertices.

We stress that the difference between Scenario 0 and Scenario I is related only to the ratio $(g(T)/\sqrt{4\pi \cdot 0.3})^6$, which changes the $T$ dependence of $\qhat(T)$ drastically. Other Scenarios II-IV include the momentum/energy dependence of the strong coupling in jet and radiative vertices, which modifies $T$ as well as $E$ dependencies of $\qhat$ and suppresses the low $T$-rise of Scenario 0. We note that Scenario II (motivated by the QLBT model) can not be associated directly with the LBT calculations \cite{He:2015pra}, since it is based on the DQPM, where only jet and radiative vertices have been replaced in line with the QLBT model. We recall that a similar finding of the strong dependence of the drag coefficient $A(T)$ of a heavy quark on the strong coupling $g$ and the parton masses has been pointed out in Ref. \cite{Berrehrah:2016led}.

As seen from Fig. \ref{fig:qhat-T_cummulative}, the spread of the different model results for $\qhat$ is very large and strongly depends on the model assumptions as well on the method used for the extraction of $\qhat$ from the heavy-ion data on the ratio $R_{AA}$ (the nuclear modification factor of high $p_T$ transverse spectra of hadrons in $A+A$ collisions versus $p+p$ collisions scaled with $N_\text{coll}$ -- the number of binary collisions in $A+A$) and flow coefficients $v_n$.

\begin{figure*}[ht!]
    \centering
    \includegraphics[width=0.8\textwidth]{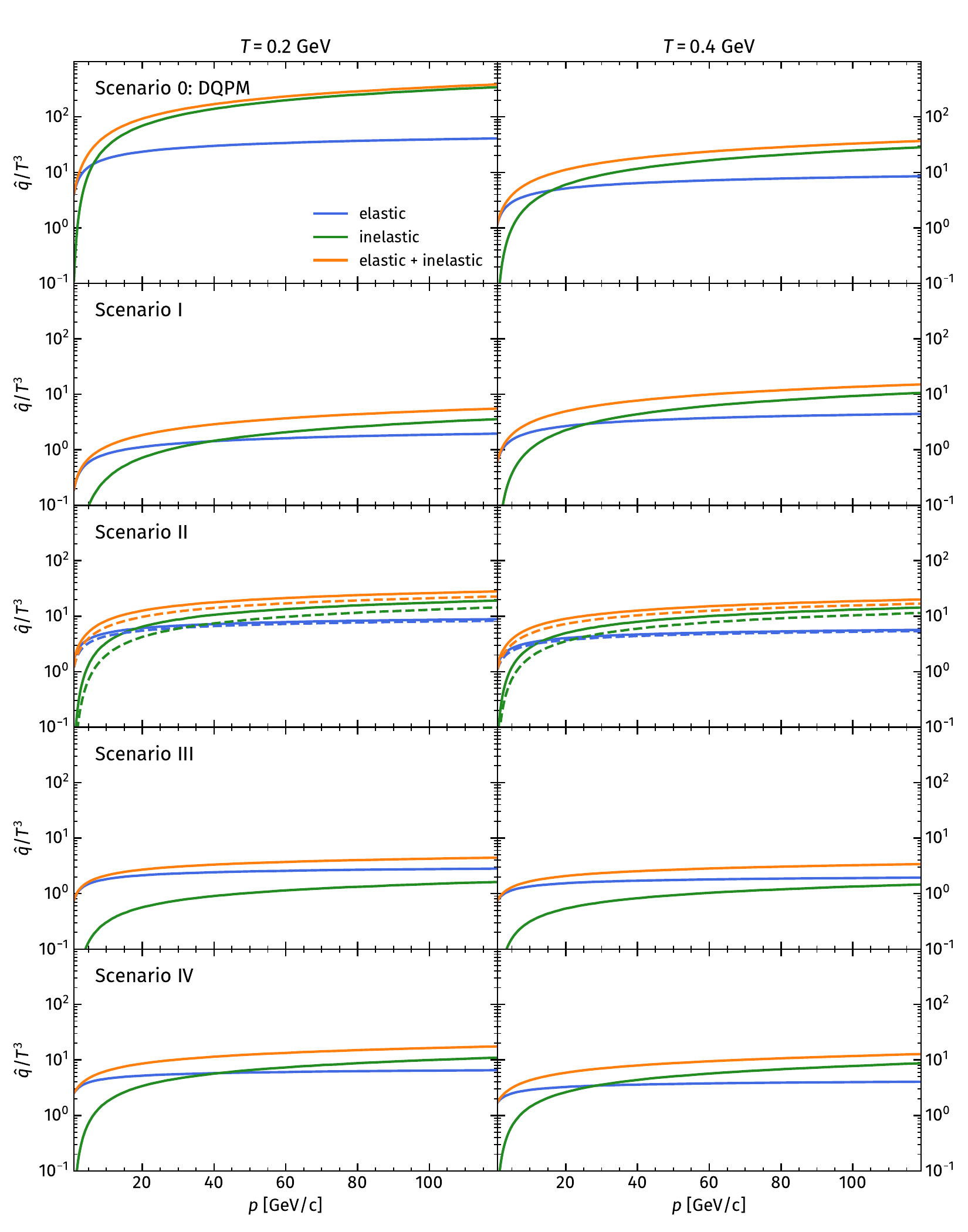}
    \caption{
        Momentum dependence of the scaled $\qhat$ coefficient for different scenarios for a jet parton with medium temperature $T=0.2$ GeV (left side) and $T=0.4$ GeV (right side). Blue lines represent results for elastic processes only and orange lines represent results for the sum of elastic and inelastic contributions.
    }
    \label{fig:qhat-p_scenarios}
\end{figure*}

\begin{figure*}[ht!]
    \centering
    \includegraphics[width=0.8\textwidth]{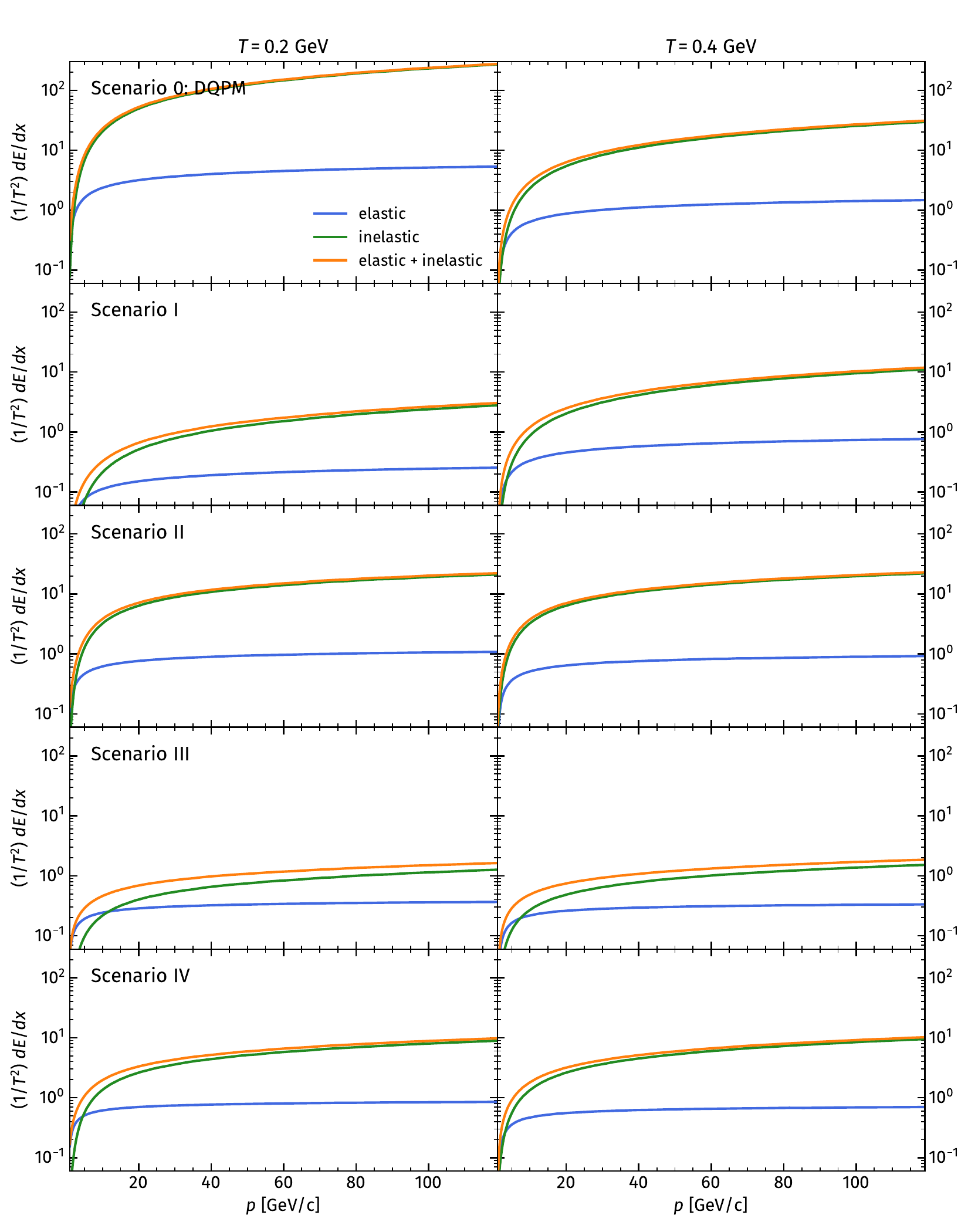}
    \caption{
        Momentum dependence of the scaled energy loss $dE/dx$ for different scenarios for a jet parton with medium temperature $T=0.2$ GeV (left side) and $T=0.4$ GeV (right side). Blue lines represent results for elastic processes only and orange lines represent results for the sum of elastic and inelastic contributions.
    }
    \label{fig:elos-p_scenarios}
\end{figure*}

In Figs. \ref{fig:qhat-p_scenarios} and \ref{fig:elos-p_scenarios} we present the jet momentum dependence of $\qhat/T^3$ and $(1/T^2) dE/dx$, respectively, for different scenarios for a jet parton at medium temperature $T=0.2$ GeV (left side) and $T=0.4$ GeV (right side). The blue lines represent results for elastic processes only, and the orange lines represent the results for the sum of elastic and inelastic contributions. Here one can see that the scaled $\qhat$ and $dE/dx$ for elastic and elastic + inelastic collisions grow with increasing of jet momentum $p$ for all scenarios for the coupling constants. The $p$ dependence of the elastic $\qhat$ and $dE/dx$ is weaker than for the inelastic one. 

As follows from Figs. \ref{fig:qhat-p_scenarios} and \ref{fig:elos-p_scenarios}, the absolute values of $\qhat/T^3$ and $dE/dx$ depend strongly on the choice of the strong coupling in the interaction vertices. The default DQPM (Scenario 0) leads to a larger transverse momentum broadening (Fig. \ref{fig:qhat-p_scenarios}) and a larger energy loss (Fig. \ref{fig:elos-p_scenarios}) at low $T$ compared to the other scenarios. This is due to the fact that the DQPM strong coupling $g$ grows rapidly at $T\to T_c$, and at low $T$ it is much larger than $g$ for the other scenarios, where its value depends on the energy/momentum transfer (cf. Fig. \ref{fig:coupling_comparison}). In other words, Scenario 0 corresponds to a fast "thermalization" of the jet parton and emitted gluon in the sQGP, since the $g$ of DQPM is evaluated for the thermal medium.

The different scenarios for the jet parton interaction in the sQPG can be tested in future studies with realistic calculations of $R_{AA}$ and $v_2$ based on the PHSD microscopic transport approach \cite{Cassing:2008nn,Moreau:2019vhw} by incorporating the radiative processes and accounting for the LPM effect. We note that the sensitivity of the $R_{AA}$ and $v_2$ to the choice of the strong coupling has been addressed within the BAMPS model by comparing the results for constant $\alpha_s=0.3$ with those for a temperature dependent $\alpha_s(T)$ \cite{Senzel:2020tgf}. This sensitivity might be larger within the DQPM based description of jet attenuation in the sQGP medium due to the additional temperature dependence of the DQPM parton masses and widths.


\section{Relation between \texorpdfstring{$\eta/s$}{eta/s} and \texorpdfstring{$T^3/\qhat$}{T**3/qhat}}
\label{sec:eta-qhat}

Finally, we explore the relation of elastic $\qhat$ to the ratio of the specific shear viscosity to entropy density $\eta/s$. It is important to recognize that the jet quenching parameter $\qhat/T^3$ serves as a direct measure of the parton coupling strength within the medium. Furthermore, generally one can anticipate that a higher coupling strength will correspond to a lower value of $\eta/s$ \cite{Majumder:2007zh}.

In Ref. \cite{Muller:2021wri} it was proposed that in the weakly coupled limit $\eta/s$ is proportional to $T^3/\qhat$:
\begin{equation}
    \eta / s \approx 1.25 \, \frac{T^3}{\qhat}.
    \label{eq:etas_qhat}
\end{equation}
This approximation can be tested within more flexible quasiparticle model frameworks, where we can explore properties of the QGP beyond the weakly interacting limit and grasp understanding towards the regime described by lQCD predictions for thermodynamic observables. The violation of the limit and more rigorous description of the QGP medium is especially relevant for moderate jet energies. It is generally expected that the assertion from Ref. \cite{Muller:2021wri}, which posits the temperature dependence of $\eta/s$ as a linearly growing function of $T^3/\qhat$, holds predominantly in the high-temperature, weak coupling regime. Conversely, for temperatures $T \le 2 T_c$, the ratio $T^3/\qhat$ is expected to decrease rapidly, significantly deviating from the linear trajectory described by $1.25 T^3/\qhat$ (for more details, see discussion in \cite{Karmakar:2023ity}). Thus, there are two regimes of a weakly and strongly coupled medium which lead to different jet attenuation in the medium. The considered quasiparticle description differs from our model, and therefore, here we aim to provide our findings employing different coupling scenarios for the thermal and jet vertices for the elastic scattering.

\begin{figure}[ht!]
    \centering
    \includegraphics[width=\columnwidth]{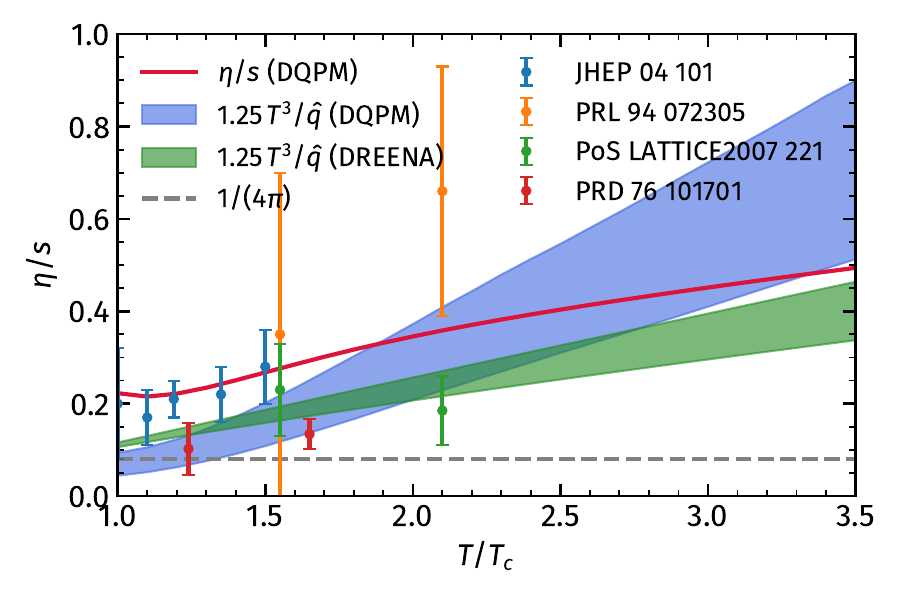}
    \includegraphics[width=\columnwidth]{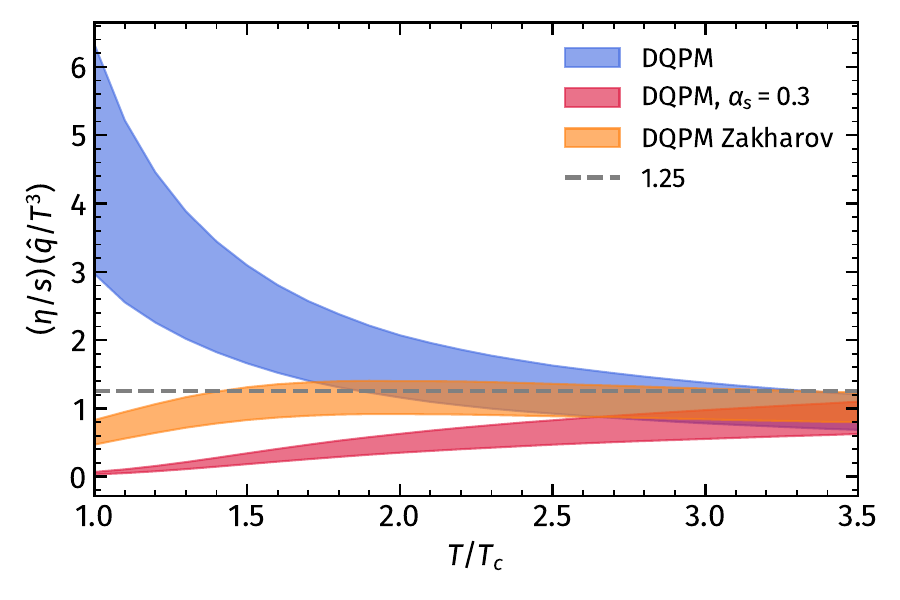}
    \caption{
        Upper plot: specific shear viscosity $\eta/s$ obtained from Eq. (\ref{eq:etas_qhat}) as a function of the scaled temperature $T/T_c$: the blue area corresponds to the Scenario 0 -- the DQPM with thermal $g(T)$; the red solid line shows $\eta/s$ computed by the RTA approach within the DQPM from Ref. \cite{Soloveva:2019xph}; the green area shows the estimates from the DREENA model \cite{Karmakar:2023ity}. The widths of the areas correspond to the jet momentum range of $3-10$ GeV$/c$. The dashed gray line shows the infinitely strong coupling limit or Kovtun-Son-Starinets bound \cite{Kovtun:2004de} $(\eta/s)_{\text{KSS}} = 1/4\pi$. The symbols display lQCD data for pure SU(3) gauge theory taken from Refs. \cite{Astrakhantsev:2017nrs} (blue circles), \cite{Nakamura:2004sy} (orange circles), \cite{Sakai:2007cm} (green circles), \cite{Meyer:2007ic} (red circles). 
        Lower plot: product of $\eta/s$ and $\qhat/T^3$ as a function of the scaled temperature $T/T_c$. The dashed gray line shows the weak interaction limit \cite{Muller:2021wri}. The blue area corresponds to the Scenario 0 (from the upper plot) -- DQPM with thermal $g(T)$, the red area shows the Scenario I -- the DQPM with $\alpha_s=0.3$; the orange area indicates the Scenario II (Zakharov coupling in jet vertex).
    } 
    \label{fig:etas}
\end{figure}

In Fig. \ref{fig:etas} we show the specific shear viscosity $\eta/s$ as a function of the scaled temperature $T/T_c$ (upper plot) and the ratio of $\eta/s$ to $T^3/\qhat$ as a function of the scaled temperature $T/T_c$ (lower plot). The red solid line in the upper plot shows $\eta/s$ computed by the RTA (relaxation-time approximation) approach within the DQPM \cite{Soloveva:2019xph} in comparison to the lQCD data (symbols) for gluodynamics ($N_f=0$). The blue area indicates the DQPM results (Scenario 0) for $1.25\, T^3/\qhat$ computed for the jet momentum range of $3-10$ GeV$/c$ (following the DREENA-A selection), which defines the width of the spread results. The DQPM result is compared to the estimates from the DREENA model \cite{Zigic:2021rku, Karmakar:2023ity} (green area). As expected, the scaled results $1.25\, T^3/\qhat$ and $\eta/s$ agree only at large $T \approx 2-3.5 T_c$, while at low $T$ the ratio $T^3/\qhat$ decreases with $T$ stronger than $\eta/s$ -- similar to the DREENA model.

The deviation from the weak-coupling scaling \eqref{eq:etas_qhat} is demonstrated in the lower part of Fig. \ref{fig:etas}, which shows the ratio of $(\eta/s)(T)$ to $T^3/\qhat$ as a function of the scaled temperature $T/T_c$. The dashed gray line corresponds to the weak interaction limit 1.25 from Ref. \cite{Muller:2021wri}. The blue area stands for the DQPM results for Scenario 0, with thermal coupling $g(T)$ for the thermal and jet vertices, which shows a strong deviation from 1.25 for lower $T$. The red area illustrates the Scenario I -- the DQPM with $\alpha_s=0.3$; the orange area stands for Scenario II (Zakharov coupling in jet vertex). These scenarios show much flatter distributions.

Thus, our results demonstrate that the ratio $\eta/s$ to $T^3/\qhat$ has a strong dependence, especially when approaching $T_c$, on the choice of the strong coupling $g$ in the scattering vertices. For the models with a temperature-dependent $g(T)$, as in the DQPM, the ratio deviates from a constant at low $T$, which corresponds to the strong interaction regime, and it approaches the scale of 1.25 at high $T$ as predicted in Ref. \cite{Muller:2021wri}, where the strong coupling $g(T)$ decreases to a small value of 0.3.

\section{Conclusions}
\label{sec:conclusion}

We have studied the jet transport coefficients, such as $\qhat$ -- the squared average transverse momentum exchange between the medium and the fast parton per unit length -- as well as $dE/dx$ -- the energy loss per unit length, including the elastic and inelastic reactions of jet partons with the thermal sQGP described by DQPM in terms of strongly interacting quasiparticles whose properties are extracted from the comparison to the lQCD equation of state. The transport coefficients have been calculated within kinetic transport theory by accounting for the independent elastic and inelastic collisions of a jet parton with a thermal parton. The contributions of elastic ($q+q \to q+q$ and $q+g \to q+g$) and inelastic ($q+q \to q+q+g$ and $q+g \to q+g+g$) reactions have been calculated by evaluating the leading-order Feynman diagrams with effective propagators and vertices from the DQPM, accounting for all channels and their interferences without approximations \cite{Grishmanovskii:2023gog}. 

We have studied different scenarios for elastic and inelastic scattering of jet parton with a parton from the sQGP medium, starting with the "default" DQPM calculations, where one assumes that all scattered partons, as well as the emitted gluon for inelastic reactions, are in thermal equilibrium. However, since the scattered jet parton and the emitted gluon might be out of equilibrium, we have additionally investigated the influence of different choices for the strong coupling constant $\alpha_s$ in thermal, jet parton, and radiative vertices, as considered in the literature.

Our findings can be summarized as following:\\
$\bullet$ We have found a strong dependence of transport coefficients on the choice of the strong coupling $g$ used in thermal, jet parton, and emitted gluon vertices (cf. Table \ref{tbl:scenarios-alphas}). This dependence is stronger for the inelastic reactions compared to elastic ones due to the extra $g^2$ in the gluon emission vertex.\\
$\bullet$ A strong rise of $\qhat$ for elastic and inelastic collisions and an increase of $dE/dx$ for inelastic reactions in the vicinity of $T \to T_c$ for the default DQPM scenario is related to the rapid rise of the DQPM coupling $g(T)$ at low $T$ -- cf. Fig. \ref{fig:coupling_comparison}. The scenario with constant $\alpha_s=0.3$ shows a decrease of transport coefficients at low $T$, which is related to the structure of the squared matrix element of the DQPM. Other scenarios with different momentum-dependent jet parton and gluon emitted vertices show a flatting of $\qhat$ and a less steep decrease of $dE/dx$ at low $T$.\\
$\bullet$ We have found a rise of $\qhat$ and $dE/dx$ with jet momentum $p$, which is stronger for inelastic reactions compared to the elastic ones.\\
$\bullet$ The contribution of inelastic reactions to the total $\qhat$ is the strongest for the default DQPM scenario due to the larger value of the strong coupling $g(T)$ compared to other scenarios.\\
$\bullet$ We have observed a strong dependence of inelastic $\qhat$ on the mass of the emitted gluon -- $\qhat$ is larger for the case with $M_g=0$ compared to the case with thermal mass of the emitted gluon $M_g=M_g(T)$ for all jet momenta $p$; $\qhat$ decreases with increasing mass of the emitted gluon; similar observations also hold for $dE/dx$.\\
$\bullet$ We have checked the validity of the scaling $\eta/s \approx 1.25 \frac{T^3}{\qhat}$, proposed in Ref. \cite{Muller:2021wri}, for the DQPM model. We have demonstrated that the ratio $\eta/s$ to $T^3/\qhat$ strongly depends on the choice of $\alpha_s$ in the scattering vertices. It rises steeply with a lowering of $T$ for the case of a temperature-dependent strong coupling $g(T)$, only at large $T$ the ratio falls to the predicted scaling value of 1.25 \cite{Muller:2021wri}, while for constant $\alpha_s$ and transverse momentum dependent $g$ the ratio approaches a constant already at $T \approx 1.5 T_c$. Thus, in line with the findings in Ref. \cite{Karmakar:2023ity} for the HTL-based model, this scaling is valid only in the weak coupling regime of large $T$ and violated for the strong coupling regime of small $T$.

Our findings are relevant for the interpretation of experimental observables on jet attenuation in the medium and extraction of the transport coefficients $\qhat$ and $dE/dx$ as well as their relation to $\eta/s$ from heavy-ion data. This will be addressed in future study within the PHSD transport approach.


\section*{ACKNOWLEDGEMENTS}

The authors acknowledge inspiring discussions with J. Aichelin, W. Cassing, M. Djordjevic, P.-B. Gossiaux, O. Kaczmarek and F. Senzel. Furthermore, we acknowledge support by the Deutsche Forschungsgemeinschaft (DFG, German Research Foundation) through the grant CRC-TR 211 "Strong-interaction matter under extreme conditions" -- project number 315477589 -- TRR 211. I.G. also acknowledges support from the "Helmholtz Graduate School for Heavy Ion Research". This work is supported by the European Union's Horizon 2020 research and innovation program under grant agreement No 824093 (STRONG-2020). The computational resources have been provided by the Goethe-HLR Center for Scientific Computing.


\bibliography{refs}

\end{document}